%%%%%%%%%%%%%%%%%%%%%%%%%%%%%%%%%%%%%%%%%%%%%%%%%%%%%%%%%%%%%%%%%
%%                                                             %% 
%%          DUALITY IN TWISTED N=4 SUPERSYMMETRIC              %% 
%%            GAUGE THEORIES IN FOUR DIMENSIONS                %%
%%                                                             %%  
%%           J.M.F. Labastida  and  Carlos Lozano              %%
%%                                                             %% 
%%                                                             %%
%%                                                             %%     
%%                       June, 1998                            %%
%%                                                             %%
%%%%%%%%%%%%%%%%%%%%%%%%%%%%%%%%%%%%%%%%%%%%%%%%%%%%%%%%%%%%%%%%%

%%
%%   PACS: 11.15.-q; 11.30.Pb; 02.40.-k
%%
%%   KEYWORDS: Supersymmetry; Topological Quantum Field Theory; Duality 
%%
%%
\input phyzzx.tex
%\input PHY.tex
%\input sphy.tex
%
%\draft
\tolerance=500000 \overfullrule=0pt

        \def\cmp{Commun. Math. Phys.}
     
    \def\jgp{J. Geom. Phys.} 
\def\jmp{J. Math. Phys.}

\def\mrl{Math. Res. Lett.}  \def\np{Nucl. Phys.}
\def\pl{Phys. Lett.}

\def\re{{\hbox{\rm Re}}}
\def\im{{\hbox{\rm Im}}}

\def\mani{\cal{M}}

\def\cn{{\cal{N}}}
\def\ad{\hbox{\rm ad}}
\def\mod{\hbox{\rm mod}}

\def\ex{{\hbox{\rm e}}}  
\def\exp{{\hbox{\rm exp}}} 
\def\tr{{\hbox{\rm Tr}}}
\def\too{\longrightarrow}
\def\half{{1\over 2}} 

\def\to{\rightarrow}

\def\sqr#1#2{{\vcenter{\vbox{\hrule height.#2pt
        \hbox{\vrule width.#2pt height#1pt \kern#1pt
           \vrule width.#2pt}
        \hrule height.#2pt}}}}

\def\deriv{{\cal D}}
%
%en modo horizontal, \sqr64\sqr64 y \hskip -5pt \hskip -7pt
%

\def\raiz{\sqrt{2}}
\def\dalpha{{\dot\alpha}}

\def\mapright#1{\smash{\mathop{\longrightarrow}\limits^{#1}}}

%
%alternativa: \buildrel ....... \over .....  ;ejemplos:
%\buildrel \alpha\beta \over \longrightarrow ,
%\buildrel \rm def \over = , etc.
%

%
%--------------- MATH LETTERS -------------------
%
\font\upright=cmu10 % scaled\magstep1
\font\sans=cmss12
\def\ssf{\sans}

\def\ZZ{\hbox{\rlap{\ssf Z}\kern 2.7pt {\ssf Z}}}
\def\zz{\relax\ifmmode\mathchoice
{\hbox{\cmss Z\kern-.4em Z}}{\hbox{\cmss Z\kern-.4em Z}}
{\lower.9pt\hbox{\cmsss Z\kern-.4em Z}}
{\lower1.2pt\hbox{\cmsss Z\kern-.4em Z}}\else{\cmss Z\kern-.4em
Z}\fi}
\def\mt{\rlap{\ssf T}\kern 3.0pt{\ssf T}}
\def\identity{{\upright\rlap{1}\kern 2.0pt 1}}
\def\inbar{\vrule height1.5ex width.4pt depth0pt}
\def\mininbar{\vrule height.75ex width.3pt depth0pt} 
\def\cc{\relax\,\hbox{$\mininbar\kern-.3em{\sevenrm C}$}}
\def\CC{\relax\,\hbox{$\inbar\kern-.3em{\rm C}$}}
\def\QQ{\relax\,\hbox{$\inbar\kern-.3em{\rm Q}$}}
\def\RR{\relax{\rm I\kern-.18em R}}
\def\PP{\relax{\rm I\kern-.18em P}}
\def\CP{\CC\PP}
\font\cmss=cmss10 \font\cmsss=cmss10 at 7pt

\def\IL{\relax{\rm I\kern-.18em L}}
\def\IH{\relax{\rm I\kern-.18em H}}
\def\IB{\relax{\rm I\kern-.18em B}}
\def\ID{\relax{\rm I\kern-.18em D}}
\def\IE{\relax{\rm I\kern-.18em E}}
\def\IF{\relax{\rm I\kern-.18em F}}
\def\IG{\relax\hbox{$\inbar\kern-.3em{\rm G}$}}
\def\IGa{\relax\hbox{${\rm I}\kern-.18em\Gamma$}}
\def\IH{\relax{\rm I\kern-.18em H}}
\def\II{\relax{\rm I\kern-.18em I}}
\def\IK{\relax{\rm I\kern-.18em K}}
\def\IP{\relax{\rm I\kern-.18em P}}
\def\IQ{\relax\hbox{$\inbar\kern-.3em{\rm Q}$}}
%%
%%-------------------------------------------------
%%
%%%%%%%%%%%%%%%%%%%%%%%%%%%%%%%%%%%%%%%%%%%%%%%%%%%%%%%%%%%%%%%%%
%%                                                             %%        
%%%%%%%%%%%%%%%%%%%%%%%%%%%%%%%%%%%%%%%%%%%%%%%%%%%%%%%%%%%%%%%%%
%%                                                                           
%%   
%%

\Pubnum={CERN-TH/98-174 \cr US-FT-8/98 \cr hep-th/9806032}
%\pubnum={CERN-TH/98-174 }
\date={June, 1998}
%\pubtype={}
\titlepage

\title{DUALITY IN TWISTED $\cn=4$ SUPERSYMMETRIC GAUGE 
THEORIES IN FOUR DIMENSIONS} 
\author{J. M. F. Labastida {\twelverm and}  
Carlos Lozano 
\foot{e-mail: lozano@fpaxp1.usc.es}}
\address{ \vbox{
\baselineskip16pt Theory Division, CERN\break CH-1211 Geneva 23,
Switzerland
\break {\rm and} \break 
Departamento de F\'\i sica de Part\'\i culas
\break Universidade de Santiago de Compostela\break 
E-15706 Santiago de Compostela, Spain}}

\abstract{\vbox{
\baselineskip16pt
We consider a twisted version of the four-dimensional 
$\cn=4$ supersymmetric Yang-Mills theory with gauge groups $SU(2)$ and $SO(3)$, and bare masses for two of its chiral multiplets, thereby breaking $\cn=4$ 
down to $\cn=2$. Using the wall-crossing technique introduced by Moore 
and Witten within the $u$-plane approach to twisted topological field theories, 
we compute the partition function and all the topological correlation functions for the case of simply-connected spin four-manifolds of simple type. 
By including 't Hooft fluxes, we analyse the properties of the resulting formulae under duality transformations. The partition function transforms in the same way as the one first presented by Vafa and Witten for another twist of the $\cn=4$ supersymmetric theory in their strong coupling test of $S$-duality. Both partition functions coincide on $K3$. The topological correlation functions turn out to transform covariantly under duality, following a simple pattern which seems to be inherent in a general type of topological quantum field theories.}}

%\vskip1cm
%\noindent CERN-TH/98-174
%\vskip-0.5cm
%\noindent May, 1998  
%\vskip-0.5cm 
%\noindent{\bf Revised version}

\endpage  
\pagenumber=1

\chapter{Introduction}
\REF\tqft{E. Witten, ``Topological Quantum 
Field Theory"\journal\cmp&117 (88)353.}
\REF\swequ{N. Seiberg and E. Witten, ``Electric-Magnetic Duality, Monopole 
Condensation, and Confinement in $\cn=2$ Supersymmetric Yang-Mills Theory", 
{\sl Nucl. Phys.} {\bf B426} (1994), 19,  
Erratum, {\sl ibid}. {\bf B430} (1994), 485; hep-th/9407087.}
\REF\swotro{N. Seiberg and E. Witten, ``Monopoles, Duality and 
Chiral Symmetry
Breaking in N=2 Supersymmetric QCD", {\sl Nucl. Phys.}
{\bf B431} (1994), 484; hep-th/9408099.}
\REF\monop{E. Witten, ``Monopoles and Four-Manifolds"
\journal\mrl&1 (94)769; hep-th/9411102.}
\REF\wimoore{G. Moore and E. Witten, ``Integration over the $u$-plane 
in Donaldson Theory", hep-th/9709193.}

During the last few years we have witnessed an important development of topological quantum field theory in four dimensions. The use, in the context 
of the Donaldson-Witten theory [\tqft], of exact results on supersymmetric gauge theories [\swequ,\swotro] led to the discovery of an important new set of topological invariants: the Seiberg-Witten invariants [\monop]. It turned out that the Donaldson invariants for four-manifolds, among other topological invariants, 
can be written in terms of these. The discovery of the Seiberg-Witten 
invariants  triggered the study of new types 
of topological quantum field theories. For some years now, 
there has been considerable 
interest in studying the twisted counterparts of several extended 
supersymmetric gauge theories, from which many astonishing links between 
the topology of low-dimensional manifolds and the dynamics of strongly-coupled 
supersymmetric gauge theories have been unveiled. 

Last year, important progress on the formulation of some topological quantum field theories in terms of effective actions associated to supersymmetric gauge theories was achieved by Moore and Witten [\wimoore]. They studied the problem of integrating over effective theories ($u$-plane integration) and introduced the technique of wall crossing to fix some of the unknown quantities which are present in the procedure. In their work they rediscovered known results for manifolds of simple type, and provided a general formulation for general manifolds and for a wide variety of topological quantum field theories related to twisted supersymmetric gauge theories with and without matter multiplets.  The aim of this paper is to apply their techniques to compute all the topological correlation functions of a topological quantum field theory based on twisted $\cn=4$ supersymmetric gauge theory with gauge groups $SU(2)$ and $SO(3)$ for the case of  simply-connected spin four-manifolds of simple type.
\REF\yamron{J. P. Yamron, ``Topological Actions in Twisted Supersymmetric
Theories"\journal\pl&B213 (88)325.} 
\REF\marcus{N. Marcus, ``The other Topological Twisting of $\cn=4$ 
Yang-Mills"\journal\np&B452 (95)331; hep-th/9506002.}
\REF\noso{J. M. F. Labastida and Carlos Lozano, ``Mathai-Quillen formulation 
of Twisted $\cn=4$ Supersymmetric Gauge Theories in Four 
Dimensions"\journal\np&B502 (97)741; hep-th/9702106.}
\REF\vafa{C. Vafa and 
E. Witten, ``A Strong Coupling Test of 
S-Duality"\journal\np&B431 (94)3; hep-th/9408074.}

$\cn=4$ supersymmetric gauge theories are conformal field theories for which 
it is believed that the duality symmetry proposed by Montonen and Olive \REF\monoli{C. Montonen and D. Olive, ``Magnetic Monopoles as Gauge Particles?"\journal\pl&B72 (77)117.}[\monoli] holds exactly. This feature is also believed to be shared by other finite field theories as, for example, 
the one which originates after introducing equal masses for two of the chiral multiplets of the $\cn=4$ supersymmetric gauge theory, thus breaking the $\cn=4$ supersymmetry down to $\cn=2$. Upon twisting, all these theories have their topological counterparts. Actually, theories with $\cn=4$ supersymmetry lead to three different topological quantum field theories [\yamron-\noso].
It is natural to expect that the topological quantum field theories which result from the twist of a theory having the duality symmetry also possess such a symmetry. This was tested by Vafa and Witten [\vafa]
for one of the twisted theories arising from $\cn=4$ supersymmetric gauge theories, for the case of gauge groups of rank $1$ and a wide varity of 
four-manifolds. In their work they used known mathematical results to verify that the partition function of the theory does indeed transform as expected under duality transformations.

In this paper we do not attempt to rediscover the partition function computed by Vafa and Witten using wall-crossing techniques within the $u$-plane approach. We will apply these techniques to compute topological correlation functions of a topological theory for which one expects the same type of duality properties as the ones found in the theory considered by Vafa and Witten. The theory is based on another twist of the $\cn=4$ supersymmetric gauge theories, also 
known as the half-twist, or twist leading to adjoint non-Abelian monopoles [\yamron,\noso]. Actually, we will be considering a more general theory in which equal masses are introduced for two of the chiral multiplets. The model possesses an important feature, which makes it more attractive than the one considered in [\vafa]: topological correlation functions different from the partition function are non-trivial. Explicit formulae for these correlation functions will teach us how correlators transform under duality in theories 
for which the duality symmetry holds exactly.

 The main goal of this paper is to compute the generating function of all the  topological correlation functions of the topological quantum field theory under consideration, for the case of simply-connected  spin manifolds of simple type. We find that the partition function possesses, for any value of the mass parameter, the same type of duality transformation properties as the one considered by Vafa and Witten, and that on $K3$ both actions coincide. Since our analysis is based on wall-crossing techniques and $u$-plane integration, the partition function, as well as the topological correlation functions, are expressed in terms of the Seiberg-Witten invariants. The duality transformations of the correlation functions turn out to be very natural. 
As we show below, they are the same as those inherent in a simple Abelian topological model.

The paper is organized as follows. In sect. 2 we review the structure of the moduli space of $\cn=4$ supersymmetric gauge theories in four dimensions. In sect. 3 we describe the topological quantum field theory involving adjoint non-Abelian monopoles, which results from one of the twists of the $\cn=4$ supersymmetric gauge theory. In sect. 4 we introduce the $u$-plane integral 
and use wall-crossing techniques to derive the contributions from the twisted 
effective theories at the singularities of the low-energy effective description, and we present the explicit form of the generating function for the topological correlation functions. In sect. 5 we study its transformation properties under duality. In sect. 6 we analyse the massless limit and the $\cn=2$ limit, which leads to Donaldson invariants. In addition, we also show in this section that the partition function coincides on $K3$ with the one found by Vafa and Witten in [\vafa], and we extend their results by presenting the full generating function of topological correlators for the massive theory. Finally, in sect. 7 we state our conclusions. An appendix deals with a set of useful identities and definitions used in the paper.

\endpage

%%%%%%%%%%%%%%%%%%%%%%%%%%%%%%%%%%%%%%%%%%%%%%%%%%%%%%%%%%%%%%%%%
%%                                                             %%
%%%%%%%%%%%%%%%%%%%%%%%%%%%%%%%%%%%%%%%%%%%%%%%%%%%%%%%%%%%%%%%%% 

\chapter{The moduli space of $\cn=4$ supersymmetric gauge theories 
in four dimensions}

In this section we  review  some aspects of the Seiberg-Witten solution 
for the low-energy effective description of the four-dimensional $\cn=4$ supersymmetric gauge theory.

\section{$\cn=4$ supersymmetric gauge theory}

We begin with several well-known remarks concerning the $\cn=4$ supersymmetric 
gauge theory on flat ${\RR}^4$.  From the point of view of $\cn=1$ superspace, 
the theory contains one $\cn=1$ vector  multiplet and three $\cn=1$ chiral 
multiplets. These 
supermultiplets are represented  in $\cn=1$ superspace  by the superfields
$V$ and $\Phi_s$ ($s=1,2,3$), which  satisfy the 
constraints $V=V^{\dag}$ and $\bar D_\dalpha \Phi_s=0$,  
$\bar D_\dalpha$ being a superspace covariant derivative\foot{We follow the 
same conventions as in [\noso].}. 
The physical component 
fields of these superfields will be denoted as follows:
$
V \rightarrow\;   A_{\alpha\dalpha},\; 
\lambda_{4\alpha},\; \bar\lambda^4{}_{\dalpha}$; 
$\Phi_s, \Phi^{\dag s} \rightarrow\; B_s,\;
\lambda_{s\alpha},\; B^{\dag s},\;
\bar\lambda^s{}_{\dalpha}.
$ 
The $\cn=4$ supersymmetry algebra has the automorphism group $SU(4)_I$, 
under which the gauge bosons are scalars, while the gauginos and the scalar fields are arranged into a pair of spinors $\lambda_{u\alpha}\oplus 
\bar\lambda^u{}_{\dalpha}$ transforming in the ${\bf 4}\oplus{\bf\bar 4}$, and a self-conjugate antisymmetric tensor $\phi_{uv}$ in the ${\bf 6}$. All the above fields take 
values in the adjoint representation 
of some compact Lie group $G$. In this paper we will work with $G= SU(2)$ 
or $SO(3)$, which are equivalent as long as we stay on $\RR^4$.  

The action takes the following form in $\cn=1$ superspace:
$$
\eqalign{
{\cal S} =& -{i\over 4\pi}\tau_0\int d^4 xd^2 \theta\, \tr (W^2) +
{i\over 4\pi}\bar\tau_0\int d^4 x d^2
\bar\theta\, \tr (W^{\dag 2}) \cr & 
+{1\over e_0^2}\sum_{s=1}^3 \int d^4 xd^2 \theta d^2
\bar\theta\, \tr(\Phi^{\dag s} \ex^V \Phi_s) \cr   
&+{i\raiz\over e_0^2}
\int d^4x d^2\theta \, \tr\bigl\{\Phi_1[\Phi_2 ,\Phi_3]\bigr\} +
{i\raiz\over e_0^2}\int d^4 xd^2\bar\theta\,\tr\bigl\{\Phi^{\dag 1} 
[\Phi^{\dag 2},\Phi^{\dag 3}]\bigr\},\cr 
}
\eqn\cuno
$$
where $W_\alpha =-{1\over 16}\bar D^2 \ex^{-V}D_\alpha \ex^V$ and 
$\tau_0={\theta_0\over{2\pi}}+{{4\pi^2 i}\over{e_0^2}}$ is the microscopic 
complexified coupling.

The theory is invariant under four independent supersymmetries which transform under $SU(4)_I$, but only one of these is manifest in the $\cn=1$ superspace formulation \cuno .
%
%$$
%\eqalign{ 
%&\delta A_{\alpha\dalpha} = -2i\bar\xi^u{}_\dalpha\lambda_{u
%\alpha}+2i\bar\lambda^u{}_\dalpha\xi_{u\alpha },\cr 
%%
%&\delta\lambda_{u\alpha} =  -iF^{+}{}_{\!\alpha}{}^{\!\beta}\xi_{u\beta}+
%i{\raiz}\bar\xi^{v\dot\alpha}\nabla_{\alpha\dot\alpha}
%\phi_{vu}-i\xi_{w\alpha}[\phi_{uv},\phi^{vw}],\cr 
%%
%&\delta\phi_{uv}={\raiz}\bigl\{\xi_u{}^\alpha\lambda_{v\alpha}
%-\xi_v{}^\alpha\lambda_{u\alpha} +
%\epsilon_{uvwz}\bar\xi^w{}_{\dalpha}\bar\lambda^{z\dalpha}\bigr\},\cr}
%\eqn\Vian
%$$
%where $F^{+}{}_{\!\alpha}{}^{\!\beta}=\sigma^{mn}{}_{\!\alpha}{}^
%{\!\beta}F_{mn}$ and $(u,v,w,z,\ldots)$ label the fundamental representation 
%${\bf 4}$ of $SU(4)_I$. In \Vian\ $\lambda_u 
%=\{\lambda_1,\lambda_2,\lambda_3,\lambda_4\}$, while  
%
%$$
%\phi_{uv}=\!\pmatrix{\!0&-\!B^{\dag 3}&\!B^{\dag 2}&-\!B_1\cr 
%                     \!B^{\dag 3}&0&\!-B^{\dag 1}&\!-B_2\cr
%                     \!-B^{\dag 2}&\!B^{\dag 1}&\!0&\!-B_3\cr
%                     \!B_1&        \!B_2&       \!B_3&\! 0\cr},\qquad
%\cases{\phi_{uv}=-\phi_{vu}, \cr
%\phi^{uv}=(\phi_{uv})^{\dag}=\phi^{*}_{vu}=
%-\half\epsilon^{uvwz}\phi_{wz}\cr}
%\eqn\Valle
%$$ 
%
The global symmetry
group of $\cn=4$ supersymmetric theories in ${\RR}^4$ is ${\cal H}=
SU(2)_L\otimes SU(2)_R\otimes SU(4)_I$, where  ${\cal K}= 
SU(2)_L\otimes SU(2)_R$ is the rotation group $SO(4)$. The fermionic  
generators of the four supersymmetries are $Q^u{}_\alpha$ and 
$\bar Q_{u\dalpha}$. They transform  as $({\bf 2},{\bf 1},{\bf \bar 4})
\oplus({\bf 1},{\bf 2},{\bf 4})$ under ${\cal H}$.

%%%%%%%%%%%%%%%%%%%%%%%%%%%%%%%%%%%%%%%%%%%%%%%%%%%%%%%%%%%%%%%%%
%%                                                             %%
%%%%%%%%%%%%%%%%%%%%%%%%%%%%%%%%%%%%%%%%%%%%%%%%%%%%%%%%%%%%%%%%% 

\section{The mass-deformed theory and the Seiberg-Witten solution}

The massless $\cn=4$ supersymmetric theory has zero beta function, and it is believed 
to be exactly finite and conformally invariant, even non-perturbatively. 
It is in fact the most promising candidate for the explicit realization 
of the strong-weak coupling duality symmetry conjectured some twenty years 
ago by Montonen and Olive [\monoli].  This theory has a moduli space of vacua in the Coulomb phase consisting of 
several equivalent copies which are interchanged by the $SU(4)_I$ symmetry. 
Each of these copies corresponds to one of the scalar fields $\phi_{uv}$  
developing a non-zero vacuum expectation value. There is a classical 
singularity at the origin of the moduli space, which is very likely to 
survive even in the quantum regime.

A more interesting theory is the one which results after deforming the $\cn=4$ supersymmetric theory by giving bare   
masses, $m\int d^4 x d^2 \theta\tr{(\Phi_1\Phi_2)}+{\hbox{\rm h.c.}}\;$,  
to two of the chiral multiplets. This mass-deformed theory still 
retains $\cn=2$ supersymmetry: the massive superfields build up an $\cn=2$ 
hypermultiplet, while the remaining chiral superfield, together with 
the vector superfield, build up an $\cn=2$ vector multiplet. The low-energy 
effective description of this theory was worked out, for the $SU(2)$ gauge group, by Seiberg and Witten in 
[\swotro]. Their results were subsequently extended to $SU(N)$ by Donagi and Witten in \REF\donagi{R. Donagi and E. Witten, ``Supersymmetric Yang-Mills 
Theory and Integrable Systems"\journal\np&B460 (96)299; 
hep-th/9510101.}[\donagi], where a  link to integrable systems was established. Some quantitative discrepancies 
between the proposed solution and 
explicit instanton calculations have been pointed out in \REF\instanton{ N. Dorey, V. V. Khoze and M. P. Mattis, 
``On Mass-Deformed $\cn=4$ Supersymmetric Yang-Mills Theory"\journal
\pl&B396 (97)141; hep-th/9612231.}[\instanton]. 
The explicit structure of the effective theory for gauge group $SU(2)$ has 
been much clarified by Ferrari \REF\ferrari{Frank Ferrari, 
``The Dyon Spectra of Finite Gauge Theories"
\journal\np&B501 (97)53; hep-th/9702166.}[\ferrari], who has 
also given a detailed account of the BPS spectrum.  

For gauge group $SU(2)$ and for generic values of the mass parameter, the moduli 
space of physically inequivalent vacua forms a one complex-dimensional compact 
manifold (the $u$-plane). This manifold parametrizes a family of elliptic curves, which encodes all the relevant information about the low-energy effective description of the theory. The explicit solution is given by the 
curve:

$$
Y^{2}=\prod_{j=1}^{3}\Bigl(X-e_{j}(\tau_0 )z-{1\over 4}e_{j}^{2}(\tau_0 ) 
m^{2}\Bigr),
\eqn\elliptic
$$
where
$$
e_1(\tau_0 )={1\over 3}(\vartheta _4^4 + \vartheta _3^4),\quad
e_2(\tau_0 )=-{1\over 3}(\vartheta _2^4 + \vartheta _3^4),\quad
e_3(\tau_0 )={1\over 3}(\vartheta _2^4 -\vartheta _4^4),
\eqn\spin
$$
and 
$\vartheta _2 (\tau )$, $\vartheta _3 (\tau )$, $\vartheta _4 (\tau )$ 
are the Jacobi theta functions -- see the appendix for more details. 

The parameter $z$ in \elliptic\ 
is a global gauge-invariant coordinate on the moduli space and it 
is a modular form of weight $2$ under the microscopic duality group. 
It differs from the physical order parameter 
$\langle \tr \, \phi ^2\rangle$ by instanton corrections [\instanton],  
which are not predicted by the Seiberg-Witten solution. The precise 
relation is given by: 
$$
z=\langle \tr\, \phi ^2\rangle - {1\over 8} m^2 e_1(\tau_0 ) +
m^2 \sum_{n=1}^{\infty}c_{n}q_0^{n},\qquad\qquad q_0=\ex^{2i\pi\tau_0}.
\eqn\instant
$$
Notice that the instanton corrections $c_{n}$ are invisible in the 
double-scaling limit $q_0\to 0$, $m\to\infty$, $4m^4 q_0=\Lambda_{0}{}^4$, 
under which the mass-deformed theory flows towards the pure gauge 
theory and $z\to u= \langle \tr\, \phi ^2\rangle$. Here $\Lambda_0$ is the dynamically generated scale of the $\cn=2$, $N_f=0$ theory. 

The low-energy description breaks down at certain points $z_i$ where the elliptic curve degenerates. This happens whenever any two of the roots of the cubic polynomial $\prod^{3}_{j=1}\left(X-e_{j}z-(1/4)e_{j}^{2} 
m^{2}\right)$ coincide. These singularities, which from the physical point of view are interpreted as due 
to BPS-saturated multiplets becoming massless, are located at the points [\swotro]:
$$
z_i= {m^2\over4} e_i
\eqn\singularity
$$

Following Ferrari [\ferrari], we choose $q_0$ small, $m$ large, with 
$m^4 q_0\sim \Lambda_{0}{}^4$.  
%This is a mild assumption that -- we expect 
%-- should not affect the final result. 
Under these circumstances, at 
strong (effective) coupling, there are two singularities at $z_2$, $z_3$, 
with $\vert z_2 - z_3\vert \sim \Lambda_0{}^2$, which flow to the 
singularities 
of the pure gauge theory in the double-scaling limit. At weak (effective) coupling, there is a third singularity, located at $z_1$, due to an electrically charged (adjoint) 
quark becoming massless. For this choice of parameters, 
we have the explicit formulas:   
$$
k^2= {\vartheta _2(\tau) ^4 \over\vartheta _3 (\tau)^4}
={\vartheta _2(\tau_0) ^4 \over\vartheta _3 (\tau_0)^4}\,
{z-z_1\over z-z_3}\raise 2pt\hbox{,}\quad
k'^2= 1-k^2={\vartheta _4 (\tau) ^4 \over\vartheta _3 (\tau)^4}
={\vartheta _4 (\tau_0) ^4 \over\vartheta _3 (\tau_0)^4}\,
{z-z_2\over z-z_3},
\eqn\modulus
$$
relating the coordinate $z$ 
to the modulus $k$ of the associated elliptic 
curve \elliptic . Here $\tau$ is the complexified effective coupling of the low-energy theory, and enters the formalism as the ratio of the two basic periods of the elliptic curve. The first period of the curve is given by the formula: 
$$
{da\over dz}= 
{\sqrt{2}\over \pi} {1\over\vartheta_{3}(\tau_0)^{2}\sqrt{z-z_{3}}}\,
K(k),
\eqn\period
$$
where 
$$
K(k)= {\pi\over2}\vartheta_3(\tau)^2
\eqn\otroperiod
$$
is the complete elliptic integral of the first kind. The second period can be computed from \period\ as ${d a_D\over d z}=\tau {d a\over d z}$. Owing to the cuts and non-trivial monodromies present on the 
$u$-plane\foot{``$z$-plane" would be more 
accurate here, but the former terminology is by now so widespread that we prefer to stick to it.}, ${d a_D\over d z}$ is not globally defined, and the actual formulas are somewhat more complicated [\ferrari]. 
In any case, the final expression for the $u$-plane integral will be invariant 
under monodromy transformations, so the above naive expression is sufficient for our purposes.

Around each of the singularities we have the following series expansion:
$$
z=z_j +\kappa_j\, q_j{}^{\half}+\cdots
\eqn\kapas
$$
where $q_j=\ex^{2\pi i\tau_j}$ is the good local coordinate at each 
singularity: $\tau_1=\tau$ for the semiclassical singularity at $z_1$, 
$\tau_2=\tau_D=-{1\over\tau}$ for the monopole singularity at $z_2$, and 
$\tau_3=\tau_d=-{1\over(\tau-1)}$ for the dyon singularity at $z_3$. 

Using \modulus , one can readily compute:
$$
\kappa_1(\tau_0)=4m^2\left({\vartheta_3\vartheta_4\over\vartheta_2}
\right)^4, \quad
\kappa_2(\tau_0)=-4m^2\left({\vartheta_2\vartheta_3\over\vartheta_4}
\right)^4,\quad
\kappa_3(\tau_0)=4m^2\left({\vartheta_2\vartheta_4\over\vartheta_3}
\right)^4.
\eqn\yespadas
$$

At the singularities, each of the periods has a finite limit when 
expressed in terms of the appropriate local coordinate:
$$
\eqalign{
\left({d a\over d z}\right)^{\,\,2}_1 &= {2\over m^2}{1\over ( 
\vartheta_3(\tau_0)\vartheta_4(\tau_0))^4}, \cr
\left({d a_D \over d z}\right)^{\,\,2}_2 &= {2\over m^2}{1\over ( 
\vartheta_2(\tau_0)\vartheta_3(\tau_0))^4}, \cr 
\left({d (a_D -a)\over d z}\right)^{\,\,2}_3 &= -{2\over m^2}{1\over ( 
\vartheta_2(\tau_0)\vartheta_4(\tau_0))^4}.
\cr}
\eqn\zhivago
$$

\endpage

%%%%%%%%%%%%%%%%%%%%%%%%%%%%%%%%%%%%%%%%%%%%%%%%%%%%%%%%%%%%%%%%%
%%                                                             %%
%%%%%%%%%%%%%%%%%%%%%%%%%%%%%%%%%%%%%%%%%%%%%%%%%%%%%%%%%%%%%%%%% 

\chapter{Twists of the $\cn=4$ supersymmetric theory}

The twisting procedure in the context of four-dimensional supersymmetric 
gauge theories was introduced by Witten in [\tqft], where he showed  
that the twisted version of the $\cn=2$ supersymmetric gauge theory with gauge 
group $SU(2)$ is a relativistic field-theory representation of the Donaldson 
theory of four-manifolds.     

In four dimensions, the global symmetry group of the extended supersymmetric 
gauge theories is of the form $SU(2)_L\otimes SU(2)_R\otimes{\cal I}$, 
where ${\cal
K}= SU(2)_L\otimes SU(2)_R$ is the rotation group, and ${\cal I}$ is the
chiral ${\cal R}$-symmetry group. The twist can be thought of either as an 
exotic realization of the global symmetry group of the theory, or as the 
gauging (with the spin connection) of a certain subgroup of the global 
${\cal R}$-current of the theory. 

While in $\cn=2$ supersymmetric gauge theories 
the ${\cal R}$-symmetry group is at most $U(2)$ and
thus the twist is essentially unique (up to an exchange of left and right), 
in the $\cn=4$ supersymmetric gauge theory the  
${\cal R}$-symmetry group is $SU(4)$ and there are three different
possibilities, each  corresponding to a different
non-equivalent embedding of the rotation group into the ${\cal R}$-symmetry
group [\yamron,\noso,\vafa].  
%% From a different perspective, it has
%% been  recently pointed out
%% [\branas] that these twisted theories appear naturally as 
%% effective
%% world-volume
%% theories on curved D-branes. 
%
\REF\wijmp{E. Witten, ``Supersymmetric Yang-Mills Theory on a Four-Manifold"
\journal\jmp&35 (94)5101; hep-th/9403195.}
\REF\masas{J. M. F. Labastida and Carlos Lozano, ``Mass Perturbations 
in Twisted $\cn=4$ Supersymmetric Gauge Theories"\journal\np&B518 (98)37; hep-th/9711132.}
\REF\coreatres{R. Dijkgraaf, J.-S. Park and B. J. Schroers, 
``$\cn=4$ Supersymmetric Yang-Mills Theory on a K{\"a}hler Surface", 
hep-th/9801066.} 
Two of these possibilities give rise to topological field theories with two
independent BRST-like topological symmetries. One of these was considered 
by Vafa and Witten [\vafa] in order to carry out an explicit test  
of $S$-duality on several four-manifolds. The key point of their calculation 
was that the partition function of this theory computes, on certain 
four-manifolds, the Euler characteristic of instanton moduli spaces, making 
it possible to fix, by comparing with known mathematical results, several unknown modular functions which could not be determined otherwise. The final computation required the introduction of a clever mass perturbation which, while breaking down the $\cn=4$ supersymmetry of the physical theory down to 
$\cn=1$, still preserves one of the topological symmetries of the theory. This  
procedure, first introduced by Witten in [\wijmp] and commonly referred to as the abstract approach, is restricted to K\"ahler manifolds with $b^{+}_2>1$.  Vafa and Witten conjectured that, in the case of the theory they were considering, the perturbation did not affect the final result for the 
partition function. Their conjecture has been recently confirmed by a careful analysis in [\masas]. Recently, Dijkgraaf et al. [\coreatres] have shown that,  on K\"ahler four-manifolds with $b^{+}_2>1$, Vafa and Witten's partition  function can be explicitly rewritten in terms of the Seiberg-Witten  invariants, thereby establishing an interesting link to four-dimensional  $\cn=2$ supersymmetric theories, which would be worthwhile to explore in the future. 

The second possibility was first addressed by Marcus [\marcus],  
and his analysis was extended in 
\REF\blauthomp{M. Blau and G. Thompson, ``Aspects of $N_{T}\geq 2$
Topological Gauge Theories and D-Branes"\journal\np&B492 (97)545; 
hep-th/9612143.}[\noso,\blauthomp]. It describes essentially intersection theory on the moduli space of complexified flat gauge connections. This 
theory was shown in [\noso] to be amphicheiral, which in this context means 
that the twisting with either $SU(2)_L$ or $SU(2)_R$ leads to the same 
result. 
\REF\corea{S. Hyun, J. Park and J.-S. Park, ``Topological QCD"\journal
\np&B453 (95)199; hep-th/9503020.}
\REF\marmon{J. M. F. Labastida and  M. Mari\~no, ``Non-Abelian 
Monopoles on Four-Manifolds"\journal\np&B448 (95)373; hep-th/9504010.}
\REF\marpol{J. M. F. Labastida and M.
Mari\~no, ``Polynomial Invariants for $SU(2)$ Monopoles"
\journal\np&B456 (95)633; hep-th/9507140.}
\REF\marth{M. Mari\~no, ``The Geometry of Supersymmetric Gauge Theories 
in Four Dimensions", Ph.D. Thesis, Universidade de Santiago de Compostela, 
October 1996; hep-th/9701128.}

The remaining possibility leads to the half-twisted theory, 
a topological theory with only one BRST symmetry [\yamron,\noso]. 
This theory is, in essence, the so-called non-Abelian monopole theory [\corea-\marth], 
the non-Abelian generalization of Witten's monopole theory 
[\monop], in the particular case in which the matter fields 
are in the adjoint representation of the gauge group. This close 
relation to twisted $\cn=2$ supersymmetric theories is not an accident, since the half-twisted theory is precisely the twisted counterpart of the 
mass-deformed $\cn=4$ supersymmetric theory. In what follows, we will 
analyse this theory in great detail.

%%%%%%%%%%%%%%%%%%%%%%%%%%%%%%%%%%%%%%%%%%%%%%%%%%%%%%%%%%%%%%%%%    
%%                                                             %%
%%%%%%%%%%%%%%%%%%%%%%%%%%%%%%%%%%%%%%%%%%%%%%%%%%%%%%%%%%%%%%%%%

\section{Adjoint non-Abelian monopoles}

The theory we wish to study can be formulated in several different -- but 
equivalent -- ways, which we now recall. It can be obtained by twisting 
the $\cn=4$ supersymmetric theory in a certain fashion first discussed by 
Yamron [\yamron]. The details are as follows. First break $SU(4)_I$ down 
to $SU(2)_A\otimes 
SU(2)_{B}\otimes U(1)$, then replace $SU(2)_L$ by its diagonal 
sum $SU(2)'_L$ with $SU(2)_{A}$. After the twisting, the symmetry 
group of the  
theory becomes  ${\cal H'} =SU(2)'_L\otimes SU(2)_R\otimes SU(2)_B
\otimes U(1)$, where the last factor is the Abelian ghost number symmetry 
of the topological theory. 
Under ${\cal H'}$, the supercharges split up as 
$$
Q^v{}_\alpha\to Q_{(\beta\alpha)},\; Q,\; Q^{i}{}_{\!\alpha}, 
\qquad \bar Q_{v\dalpha}\to \bar Q_{\alpha\dot\alpha},\;
\bar Q_{i\dot\alpha}
\eqn\elven
$$
where $i$ is an $SU(2)_B$ index. The twist gives rise to a scalar 
supercharge $Q$, which is a certain linear combination of the original 
supercharges.
%$Q = - Q^{v=2}{}_{\alpha =2} -Q^{v=1}{}_{\alpha=1}$.

The fields of the $\cn=4$ supersymmetry multiplet 
transform under ${\cal H'}$ as follows -- in the 
notation of [\yamron]:
$$
\eqalign{
A_{\alpha\dalpha} &\too A^{(0)}_{\alpha\dalpha},\cr
\lambda_{v\alpha}&\too
\chi^{(-1)}_{\beta\alpha},~
\eta^{(-1)},~\lambda^{(+1)}_{i\alpha},\cr}
\qquad\qquad
\eqalign{
\bar\lambda^v{}_{\dalpha} &\too
\psi^{(+1)}_{\alpha\dalpha},~\zeta^{(-1)}_{i\dalpha}\cr
\phi_{uv}&\too \lambda^{(-2)},~\phi^{(+2)},~G^{(0)}_{i\alpha}.
\cr}
\eqn\poocho
$$
The superscript stands for the ghost number carried by each of the fields. 
Notice that the field $\chi_{\alpha\beta}$ is 
symmetric in its spinor indices and can therefore be regarded as a 
self-dual two-form. 
As explained in [\noso], the isospin group is not manifest in the field-theory realization we are interested in. Instead, we reorganize the $SU(2)_B$ doublets 
in \poocho\ into three pairs of complex-conjugate spinors: $\lambda_{i\alpha}\to \mu_\alpha,~\bar\mu_\alpha$; $\zeta_{i\dalpha}\to \nu_\dalpha,~\bar\nu_\dalpha$; $G_{i\alpha}\to M_\alpha,~\overline M_\alpha$. Taking this into account, the 
field content of the twisted theory consists of two scalar fields $\{\phi^{(+2)},\lambda^{(-2)}\}$, two left-handed spinors 
$\{M^{(0)}_\alpha,\bar M^{(0)}_\alpha\}$,   
two auxiliary right-handed spinors $\{h^{(0)}_\dalpha,
\bar h^{(0)}_\dalpha\}$, a one-form $A_{\alpha\dalpha}^{(0)}$,  
and a self-dual auxiliary two-form  $H^{(0)}_{\alpha\beta}$ 
on the bosonic (commuting) side; and 
a scalar field $\eta^{(-1)}$, a pair of left-handed spinors $\{\mu^{(+1)}_\alpha,\bar\mu^{(+1)}_\alpha\}$,   
two right-handed spinors $\{\nu^{(-1)}_\dalpha,\bar 
\nu^{(-1)}_\dalpha\}$, a one-form $\psi_{\alpha\dalpha}^{(1)}$ 
and a self-dual two-form  $\chi^{(-1)}_{\alpha\beta}$  
on the fermionic (anticommuting) side. 

The twisted $\cn=4$ supersymmetric action breaks up into a $Q$-exact piece (that is, a 
piece which can be written as $\{Q,{\cal T}\}$, where ${\cal T}$ is a functional 
of the fields of the theory), plus a topological term proportional to 
the instanton number of the gauge configuration,
$$
{\cal S}_{\hbox{\sevenrm twisted}}= \{Q,{\cal T}\}-2\pi in\tau_{0},
\eqn\ramala
$$
with $n={1\over 16\pi^2}\int_X \tr\,(F\wedge F)={1\over 32\pi^2}\int_X \sqrt{g}\tr\,(*F_{\mu\nu}F^{\mu\nu})$, the instanton number, which is an 
integer for $SU(2)$ bundles but a half-integer for non-trivial $SO(3)$ 
bundles on spin four-manifolds. Therefore, as pointed out in [\vafa], one would expect the $SU(2)$ theory to be invariant under $\tau_0\to\tau_0+1$, while the $SO(3)$ theory should be only invariant under $\tau_0\to\tau_0+2$ on spin manifolds. Notice that, owing to \ramala, the partition function depends on 
the microscopic couplings $e_{0}$ 
and $\theta_{0}$ only through the combination $2\pi in\tau_{0}$, 
and in particular this dependence is a priori holomorphic (if we reversed  
the orientation of the manifold $X$, the partition function would depend 
anti-holomorphically on $\tau_0$). However there could be manifolds in which, 
because of some sort of holomorphic anomaly, the partition function would acquire an explicit anomalous dependence on $\bar\tau_{0}$. This seems to be the case, for example, for the partition function of the Vafa-Witten twist on $\CP^{2}$ [\vafa]. 

 In the twisting procedure, one couples the twisted action \ramala\ to arbitrary gravitational backgrounds, so as to deal with its formulation for 
a wide variety of manifolds. In general, the procedure involves the covariantization of the flat-space action, as well as the addition of curvature terms to render the new action as a $Q$-exact piece plus a topological term as in \ramala. Actually, on curved space one might think of additional topological terms -- such as 
$\int R\wedge R$ or $\int R\wedge*R$, with $R$ the curvature two-form of the manifold -- besides the one already present in \ramala. Thus, the action which comes out of the twisting procedure is not unique (even modulo $Q$-exact terms),  since it is always possible to add $c$-number terms, which vanish on flat space but are nevertheless topological. In a topological field theory in four dimensions, those terms are proportional to the Euler number $\chi$ and the signature $\sigma$ of the manifold $X$. In order to keep the holomorphicity in $\tau_0$, the proportionality constants must be functions of $\tau_0$. At this stage one does not know which particular functions to take, but clearly good transformation properties under duality could be spoiled if one does not make the right choice. It seems therefore that there exists a preferred choice of those terms, which is compatible with duality. This issue was treated in detail in [\vafa],  where it was shown that a $c$-number of the form $-i\pi \tau_0 \chi/6$ was needed in the topological action in order to have a theory with good transformation properties under duality. For the theory considered in this paper, it turns out that the $c$-number which must be present in the action has the form $-i\pi\tau_0(\chi+\sigma)/2$ if $2\chi+3\sigma=0~\mod~32$, 
and $i\pi(2\chi+3\sigma)/8$ otherwise. This will be shown in section 4. 

\REF\laplata{J. M. F. Labastida and Carlos Lozano, ``Lectures on Topological 
Quantum Field Theory", in Proceedings of the CERN-Santiago de 
Compostela-La Plata Meeting on ``Trends in Theoretical Physics", 
H. Falomir, R. Gamboa, 
F. Schaposnik, eds. (American Institute of Physics, New York, 1998); 
hep-th/9709192.}
\REF\phyrep{D. Birmingham, M. Blau, M. Rakowski and G.
Thompson, ``Topological Field Theories", {\sl Phys. Rep.} 
{\bf 209} (1991), 129.}
\REF\moore{S. Cordes, G. Moore and S. Rangoolam, ``Lectures on 2D 
Yang-Mills Theory, Equivariant Cohomology and Topological Field Theory", 
{\sl Nucl. Phys. Proc. Suppl.} {\bf 41} (1995), 184; hep-th/9411210.}
\REF\thompson{M. Blau and G. Thompson, ``Localization and Diagonalization: a
Review of Functional Integral Techniques for Low-Dimensional Gauge Theories 
and Topological Field Theories"\journal\jmp&36 (93)2192; hep-th/9501075.}
\REF\puri{J.M.F. Labastida and M. Mari\~no, ``Duality and Topological Quantum Field Theory", talk given at the ``Workshop on Frontiers of Field Theory, Quantum Gravity and String Theory", Puri, India, 1996; hep-th/9704032.}

The significance of the above field spectrum, as well as the underlying 
geometric structure of the topological theory, can be most transparently 
understood within the framework of the Mathai-Quillen formalism 
\REF\atiy{M. F. Atiyah and L. Jeffrey, ``Topological Lagrangians 
and Cohomology"\journal\jgp&7 (90)119.}[\atiy] 
(for a review of the Mathai-Quillen formalism in the context of
topological field theories, see [\laplata-\puri]). 
From this viewpoint, the theory is defined in terms of the monopole equations:
$$
\cases{ F^{+}_{\alpha\beta}+2[\overline M_{(\alpha},M_{\beta)}]=0,\cr
\deriv_{\alpha\dalpha}M^{\alpha}=0,\cr
}
\eqn\masecuaciones
$$
which characterize the fixed points of the BRST symmetry generated by $Q$.  
These equations are now interpreted as defining a section 
$s:{\cal M}\to {\cal V}$ in the trivial vector bundle  
${\cal V}=\mani\times{\cal F}$, where ${\cal M}={\cal A}\times 
\Gamma(X,S^{+}\otimes\ad P)$ is the field space, and the fibre is 
${\cal F}=  \Omega^{2,+}(X,\ad P)\oplus\Gamma (X,S^{-}\otimes\ad
P)$, whose zero locus 
-- modded out by the gauge symmetry -- is precisely the desired moduli space. 
${\cal A}$ denotes the space of connections on a principal $G$-bundle $P\to
X$, $\Gamma(X,S^{+}\otimes\ad P)$ is the space of sections of the product 
bundle $S^{+}\otimes\ad P$, that is, positive chirality spinors taking values 
in the Lie algebra of the gauge group, while $\Omega^{2,+}(X,\ad P)$ denotes 
the space of self-dual two-forms on $X$ taking values in the Lie
algebra of $G$. $\ad P$ denotes the adjoint bundle of $P$, $P\times_{\ad} 
{\bf g}$ (${\bf g}$ stands for the Lie algebra of $G$). The space of sections
of this bundle, $\Omega^0(X,\ad P)$, is the Lie algebra of the group ${\cal G}$ 
of gauge transformations (vertical automorphisms) of the bundle $P$.  

In this setting $A$ and $M_\alpha$ define the field space; $\psi$ and $\mu$ 
are ghosts living in the (co)tangent space $T^*{\cal M}$; 
$\chi^{+}$ and $\nu$ are fibre antighosts associated to the equations  \masecuaciones   
, while 
$H^{+}$ and $h$ are their corresponding auxiliary fields; finally, $\phi$ -- or 
rather its vacuum expectation value $\langle\phi\rangle$ -- gives the curvature 
of the principal 
${\cal G}$-bundle ${\cal M}\to {\cal M}/{\cal G}$, while $\lambda$ and $\eta$ 
enforce the horizontal projection ${\cal M}\to {\cal M}/{\cal G}$. The BRST 
symmetry generated by $Q$ is the Cartan model representative of the ${\cal G}$-equivariant differential on ${\cal V}$, 
while the ghost number is just a form degree. The exponential of the action of 
the theory gives, when integrated over the antighosts and their auxiliary
fields, the Mathai-Quillen representative for the Thom form of the principal 
bundle ${\cal M}\times{\cal F}\to {\cal E}={\cal M}\times_{\cal G} {\cal F}$.  

Notice that the twisted theory contains several spinor fields. This means that the theory is not well defined on those manifolds that do not admit a spin structure. As explained in 
\REF\baryon{J. M. F. Labastida and M. Mari\~no, ``Twisted Baryon 
Number in $\cn=2$ Supersymmetric QCD"\journal\pl&B400 (97)
323; hep-th/9702054.}
[\corea,\marmon,\baryon] -- see [\monop] for a related discussion --, one could try to avoid this problem by coupling the spinor fields to a fixed (background) Spin$_{c}$ structure. We will not follow this path here, and therefore we will take $X$ to be an -- otherwise arbitrary -- spin four-manifold.  

It would be interesting to know whether the mass-deformed theory has an analogue twisted version which is also a topological quantum field theory, but now with massive fields. It turns out that this is indeed the case (it was shown for the $\cn=2$ supersymmetric theory with one hypermultiplet in the fundamental representation in \REF\kungfu{M. Alvarez and J.M.F. Labastida, ``Topological 
Matter in Four Dimensions"\journal\np&B437 (95)356; hep-th/9404115.}
\REF\koreas{S. Hyun, J. Park and J.-S. Park, ``$\cn=2$ Supersymmetric QCD 
and Four-Manifolds; (I) the Donaldson and the Seiberg-Witten Invariants", 
hep-th/9508162.}[\kungfu,\koreas], and it is straightforward to extend their result to the present situation). As shown in 
\REF\zzeta{J. M. F. Labastida and M. Mari\~no, ``Twisted $\cn=2$ 
Supersymmetry with Central Charge and Equivariant Cohomology"
\journal\cmp&185 (97)37; hep-th/9603169.}[\zzeta], the theory is in fact 
an equivariant extension of the massless theory with respect to a $U(1)$ symmetry which rotates the monopole fields, $M\to \ex^{i\alpha}M$, 
$\overline M\to \ex^{-i\alpha}\overline M$. From this viewpoint, $m$ can be thought of 
as the generator of this $U(1)$ symmetry.  

The linearization of eqs. \masecuaciones\ provides a map 
$ds:{\cal M}\to {\cal F}$, which fits into the deformation complex [\noso]:
$$
\eqalign{  0&\too\Omega^0(X,\ad P)\mapright{{\cal C}}\Omega^1(X,\ad
P)\oplus\Gamma(X,S^{+}\otimes\ad P)\cr &\mapright{ds}\Omega^{2,+}(X,\ad
P)\oplus\Gamma(X,S^{-}\otimes\ad P),\cr}
\eqn\complex
$$  
where the map ${\cal C}:\Omega^0(X,\ad P)\too T{\cal M}$ is given by:
$$  
{\cal C}(\phi)=(d_A \phi,i[M,\phi]),\quad \phi\in\Omega^0(X,\ad P). 
\eqn\barbarella
$$ 
 
The index of the complex \complex\ gives the virtual dimension of the 
moduli space. One can show in this way that, for gauge group $SU(2)$:
$$
{\hbox{\rm dim}}({\cal M})= -{3\over4}(2\chi+3\sigma), 
\eqn\indices
$$ 
where $\chi$ is the Euler characteristic of the four-manifold $X$ and 
$\sigma$ its signature. Notice that the dimension of the moduli space 
does not depend on the instanton number of the gauge configuration.  

Topological invariants are obtained by considering the vacuum 
expectation value of arbitrary products of observables, which 
are operators that are $Q$-invariant but not $Q$-exact. As discussed in [\noso,\marpol], the relevant observables for this theory  
and gauge group $SU(2)$ or $SO(3)$, are precisely the same as in the Donaldson-Witten theory [\tqft]:
$$
\eqalign{
W_0 =& {1\over 8\pi^2}\tr(\phi^2), \cr
W_2 =& {1\over 8\pi^2}\tr(2\phi F+\psi \wedge \psi), \cr}
\qquad\qquad
\eqalign{
W_1 =& {1\over4\pi^2}\tr(\phi\psi), \cr
W_3 =& {1\over4\pi^2}\tr(\psi \wedge  F). \cr}
\eqn\guta
$$
The operators $W_i$ have positive ghost numbers given by $4-i$ 
and satisfy the descent equations 
$$
[Q, W_i\} = d W_{i-1},
\eqn\descent
$$
which imply that
$$
{\cal O}^{(\gamma_j)} = \int_{\gamma_j} W_j,
\eqn\noguta
$$
$\gamma_j$ being homology cycles of $X$, are observables. 

The vacuum expectation value of an arbitrary product of observables has 
the general form (modulo a term which involves the exponential of a linear combination of $\chi$ and $\sigma$),
$$
\left\langle \prod_{\gamma_j} {\cal O}^{(\gamma_j)}\right\rangle
= \sum_{n} 
\left\langle \prod_{\gamma_j} {\cal O}^{(\gamma_j)}\right\rangle_n
\ex^{-2\pi i n \tau_0},
\eqn\pumba
$$
where $n$ is the instanton number and 
$\left\langle \prod_{\gamma_j} {\cal O}^{(\gamma_j)}\right\rangle_n$
is the vacuum expectation value computed 
at a fixed value of $n$. These quantities are independent of the 
coupling constant $e_0$. When
analysed in the weak coupling limit, the contributions to the functional
integral come from field configurations which are solutions to eqs.  \masecuaciones . All the dependence of the observables on $\tau_0$ is
contained in the phases $\exp(-2\pi i n \tau_0)$ in \pumba. The question therefore arises as to whether
the vacuum expectation values of these observables have good modular
properties under $Sl(2,{\ZZ})$ transformations acting on $\tau_0$. 
One of the aims of this paper is to show that this is indeed the case,  
at least for  spin four-manifolds of simple type 
(although one could easily extend the arguments presented here to all 
simply-connected spin manifolds with $b^+_2>1$).

The ghost-number anomaly of the theory restricts the possible non-trivial topological invariants to be those for which the overall ghost number of the operator insertions matches the anomaly $-(3/4)(2\chi+3\sigma)$. Notice that since any arbitrary product of observables has necessarily positive ghost number, there will be no non-trivial topological invariant for those manifolds for which $2\chi+3\sigma$ is strictly positive. On the other hand, 
if $2\chi+3\sigma<0$, there is only a finite number -- if any -- of non-trivial topological invariants. Finally, when $2\chi+3\sigma=0$, as is the case for $K3$, for example, the only non-trivial topological invariant is the partition function. Moreover, as the physical and twisted theories are actually the same on hyper-K\"ahler manifolds as $K3$, this partition function should coincide with the one computed by Vafa and Witten for another twist of the $\cn=4$ supersymmetric theory in [\vafa]. Below we will show that this is indeed the case. Notice that this assertion does not apply to the twist first considered by Marcus, as it actually involves two independent twists, one on each of the $SU(2)$ 
factors of the holonomy group of the manifold [\yamron-\noso].   

The selection rule on the topological invariants that we have just 
discussed does not apply of course to the massive theory, as the 
mass terms explicitly break the ghost number symmetry. However, the 
following is a useful constraint. The mass perturbation $m\int d^2\theta \tr(\Phi_1\Phi_2)+\overline{m}\int d^2\bar\theta \tr(\Phi^{\dag}_1
\Phi^{\dag}_2)$ can be twisted as well, and provides Lorentz-invariant 
mass terms for the monopoles $M_\alpha$ and their partners $\mu_\alpha$ and  
$\nu_\dalpha$. These mass terms break up into a $Q$-exact piece (which can therefore be discarded when computing vevs of $Q$-invariant operators), and 
a second piece, $m\Delta {\cal L}$, which is linear in $m$ and does not depend on $\overline{m}$ -- see eqs. (4.39)-(40) in [\zzeta]. Here $\Delta {\cal L}$ is a polynomial function in the fields with net ghost number $-2$. The partition function 
of the massive theory can be interpreted as the vev of the operator $\ex^{m\Delta {\cal L}}$ in the massless theory. This vev must be 
understood as a series expansion in powers of $m$, 
$$
\left\langle\ex^{m\Delta {\cal L}}\right\rangle = \sum_{\ell=0}^{\infty} 
{m^{\ell}\over \ell!} \left\langle\left(\Delta {\cal L}\right)^{\ell}\right\rangle.
\eqn\cocoon
$$ 
Each term in the above expansion has successively lower ghost number $-2\ell$. 
As explained above, the only non-vanishing correlator will be that for which 
the net ghost number of the operator insertion ($-2\ell$ for the $\ell$-th term) equals the anomaly of the theory, $-(3/4)(2\chi+3\sigma)$. This forces the dependence of the partition function on $m$ to be of the form:
$$
\langle 1 \rangle_{m}= m^{{3\over8}(2\chi+3\sigma)} F(\tau_0,\chi,\sigma), 
\eqn\coop
$$
where $F$ is a certain function to be determined below. 

A simple check to test that \coop\ leads to the correct dependence on the mass is the following. The action \ramala, perturbed by mass terms as above, is invariant under a $U(1)$ chiral transformation -- that is, a ghost number transformation -- if the field transformations are accompanied by an appropriate rescaling of the mass parameter. However, the partition function, as well as the correlation functions, is not invariant -- unless the appropriate selection rule is satisfied -- due to a contribution from the measure that is proportional to the chiral anomaly, which is given precisely 
by \indices. Thus, these quantities should transform into themselves times a term proportional to the exponential of the chiral anomaly. This is in fact 
the behaviour of \coop\ under a rescaling of $m$.
  
\endpage

%%%%%%%%%%%%%%%%%%%%%%%%%%%%%%%%%%%%%%%%%%%%%%%%%%%%%%%%%%%%%%%%% 
%%                                                             %%
%%%%%%%%%%%%%%%%%%%%%%%%%%%%%%%%%%%%%%%%%%%%%%%%%%%%%%%%%%%%%%%%%

\chapter{Integrating over the $u$-plane}

The functional-integral (or microscopic) approach to twisted supersymmetric 
quantum field theories gives great insight into their geometric structure, 
but it does not allow us to make explicit calculations. Once 
the relevant set of field configurations (moduli space) on which 
the functional-integral is supported has been identified, the computation of 
the partition function or, more generally, of the topological correlation 
functions, is reduced to a finite-dimensional integration over the quantum 
fluctuations (zero modes) tangent to the moduli space. For this to produce  
sensible topological information, it is necessary to give a suitable 
prescription for the integration, and a convenient compactification of the 
moduli space is usually needed as well. This requires an extra input of information which, in most of the cases is at the heart of the subtle topological information expected to capture with the invariants 
themselves. 

The strategy to circumvent these problems and extract concrete predictions 
rests in taking advantage of the crucial fact that, by construction, the generating functional for topological correlation functions in a topological quantum field theory is independent of the metric on the manifold.
This implies that, in principle, these correlation functions can be computed 
in either the ultraviolet (short-distance) or infrared (long-distance) 
limits. The naive functional-integral approach focuses on the short-distance 
regime, while long-distance computations require a precise knowledge of 
the vacuum structure and low-energy dynamics of the physical theory.    

Following this line of reasoning, it was proposed by Witten in [\monop] 
that the explicit solution for the low-energy effective descriptions for a family of four-dimensional $\cn=2$ supersymmetric field theories presented in [\swequ,\swotro] could be used to perform an alternative long-distance computation of the topological correlators which relies completely on the 
properties of the physical theory. This idea is at the heart of the 
successful reformulation of the Donaldson invariants, for a certain subset of four-manifolds, in terms of the by now well-known Seiberg-Witten invariants, which are essentially the partition functions of the twisted effective Abelian theories at the singular points on the moduli space of vacua of the physical, $\cn=2$ supersymmetric theories. The same idea has been subsequently applied 
to some other theories 
\REF\highrank{M. Mari\~no and G. Moore, ``The Donaldson-Witten 
Function for Gauge Groups of Rank Larger than One",  hep-th/9802185.}
\REF\nonsimply{M. Mari\~no and G. Moore, ``Donaldson Invariants 
for Nonsimply Connected Manifolds",  hep-th/9804104.}
\REF\sanson{A. Losev, N. Nekrasov and S. Shatashvili, 
``Issues in Topological Gauge Theory", hep-th/9711108; ``Testing 
Seiberg-Witten Solution", hep-th/9801061.}[\wimoore,\marpol,\highrank-\sanson], thereby providing a whole 
set of predictions which should be tested against explicit mathematical results. The moral of these computations is that the duality structure of extended supersymmetric theories automatically incorporates, in an as yet  
not fully understood way, a consistent compactification scheme for the 
moduli space of their twisted counterparts.    

The standard computation of this sort involves an integration 
over the moduli space of vacua (the $u$-plane) of the physical theory. At 
a generic vacuum, the only contribution comes from a twisted $\cn=2$ Abelian vector multiplet. The effect of the massive modes is contained in appropriate measure factors, which also incorporate the coupling 
to gravity -- these measure factors were derived in 
\REF\sdual{E. Witten, ``On S-Duality in Abelian Gauge Theory", {\sl 
Selecta Mathematica} {\bf 1} (1995), 383; hep-th/9505186.}
[\sdual] by demanding 
that they reproduce the gravitational anomalies of the massive fields --, 
and in contact terms among the observables -- the contact term for the two-observable for the $SU(2)$ theory was derived in [\wimoore] and was subsequently extended to more general observables and other gauge groups 
in \REF\strings{M. Mari\~no and G. Moore, ``Integrating over 
the Coulomb Branch in $\cn=2$ Gauge Theory", hep-th/9712062.}
[\highrank,\sanson,\strings]. 

The total contribution to the generating function thus consists of an 
integration 
over the moduli space with the singularities removed -- which is 
non-vanishing for $b^{+}_2(X)=1$ [\wimoore] only -- plus a discrete sum 
over the contributions of the twisted effective theories at each of the 
singularities. The effective theory at a given singularity should contain, 
together with the appropriate dual photon multiplet, several charged 
hypermultiplets, which correspond to the states becoming massless at 
the singularity. The complete effective action for these 
massless states contains as well certain measure factors and 
contact terms among the observables, which reproduce the 
effect of the massive states which have 
been integrated out. However, it is not possible to fix these a priori 
unknown functions by anomaly considerations only. As first proposed in 
[\wimoore] -- see [\highrank,\nonsimply,\strings] for more 
details and further 
extensions --, the alternative strategy takes advantage of the 
wall-crossing properties of the $u$-plane integral. It was shown in [\wimoore] 
that at those points on the $u$-plane where the 
(imaginary part of the) effective coupling 
diverges, the integral has a discontinuous variation when the self-dual 
two-form $\omega$, which gives a basis for $H^{2,+}(X)$, is such that, for a given gauge configuration $\lambda\in H^{2}(X;\ZZ)$, the period 
$\omega\cdot\lambda$ changes sign. This is commonly referred to as 
``wall crossing". The points where wall crossing can take place are the singularities of the moduli space due to charged matter multiplets becoming massless -- the appropriate local effective 
coupling $\tau$ diverges there -- and, in the case of the 
asymptotically free theories, the point at infinity, $u\to\infty$, 
where also the effective electric coupling diverges owing to asymptotic 
freedom.

On the other hand, the final expression for the invariants can 
exhibit a wall-crossing behaviour at most at $u\to\infty$, so the 
contribution to  wall crossing from the integral at the 
singularities at finite values of $u$ 
must cancel against the contributions coming from the effective theories 
there, which also display wall-crossing discontinuities.  
As shown in [\wimoore], this cancellation fixes almost completely the 
unknown functions in the contributions to the topological correlation 
functions from the singularities.

\vskip2cm

%%%%%%%%%%%%%%%%%%%%%%%%%%%%%%%%%%%%%%%%%%%%%%%%%%%%%%%%%%%%%%%%%    
%%                                                             %%      
%%%%%%%%%%%%%%%%%%%%%%%%%%%%%%%%%%%%%%%%%%%%%%%%%%%%%%%%%%%%%%%%%

\section{The integral for $\cn=4$ supersymmetry}

The complete expression for the $u$-plane integral for the gauge group 
$SU(2)$ and $N_f\leq 4$ was worked out in [\wimoore]. The appropriate 
general formulas for the contact terms can be found in 
[\highrank,\sanson,\strings]. These formulas follow the conventions in [\swotro], 
according to which, for $N_f=0$, the $u$-plane is the modular curve of $\Gamma^0(4)$. In this formalism, the monodromy associated to a 
single matter multiplet 
becoming massless is conjugated to $T$. As for the $\cn=4$ supersymmetric  theory, it is 
more convenient to use instead a formalism related to $\Gamma(2)$, in 
which the basic monodromies are conjugated to $T^2$. Our formulas follow straightforwardly from those in [\wimoore,\highrank,\strings], with some 
minor changes to conform to our conventions.     

The integral in the $\cn=4$ supersymmetric case, for gauge groups $SU(2)$ or $SO(3)$ and on 
simply-connected four-manifolds, is given by the formula: 

$$
\left\langle \ex^{p{\cal O}+ I(S)}\right\rangle_{\xi}\bigg\vert_{\hbox
{\sevenrm plane}}= 
Z_u(p,S,m,\tau_0) = {2\over i}\int_{\CC}
{dz  d\bar z \over  {y}^{1/2}}
\mu(\tau) \ex^{2 p z + S^2 \hat T(z)} \Psi ,
\eqn\integral
$$
where $y=\im\,\tau$. The expression \integral\ gives the generating function for the vacuum expectation values of two of the observables in \noguta: 
$$
\eqalign{
{\cal O}&= {1\over 8\pi^2}\tr(\phi^2),\cr 
I(S)&= \int_{S}{1\over 8\pi^2}\tr(2\phi F+\psi \wedge \psi).\cr
}
\eqn\belen
$$
Here, $S$ is a two-cycle on $X$ given by the formal sum $S=\sum_{a}
\alpha_{a}S_{a}$, where $\{S_{a}\}_{a=1,\ldots,b_2(X)}$ are two-cycles representing a basis of $H_2(X)$, and $S^2\equiv \sum_{a,b}\alpha_{a}
\alpha_{b}\sharp(S_{a}\cap S_{b})$, where $\sharp(S_{a}\cap S_{b})$ 
is the intersection number of $S_{a}$ and $S_{b}$. Notice that since we 
are restricting ourselves to simply-connected four-manifolds, there is no 
non-trivial contribution from the one- and three-observables $W_{1}$ and $W_3$ in \noguta. 
The generalization to the non-simply-connected case was outlined in [\wimoore], 
and it has been recently put on a more rigorous basis in [\nonsimply].  

The measure factor $\mu(\tau)$ is given by the expression:    
$$
\mu(\tau)= f(m,\chi,\sigma,\tau_0) { d \bar \tau \over  d \bar z}
\left({d a \over  dz } \right)^{1- \half \chi  } \Delta^{\sigma/8},
\eqn\measure
$$ 
where $\Delta$ is the square root of the discriminant of the Seiberg-Witten 
curve \elliptic:
$$
\eqalign{
\Delta &= \eta(\tau_0)^{12}(z-z_1)(z-z_2)(z-z_3) = {\eta(\tau)^{12}\over 
2^3 (d a/d z)^6}
\cr
&= -{\eta(\tau_D)^{12}\over 2^3 (d a_D/d z)^6} = 
{\eta(\tau_d)^{12}\over 2^3 (d (a_D-a)/d z)^6},
\cr}
\eqn\discriminante
$$
where $\eta(\tau)$ is the Dedekind function  
and $f(m,\chi,\sigma,\tau_0)$ is a universal normalization factor which cannot be fixed a priori. It can be fixed in the $N_{f}=0$ case by comparing with  known 
results for the Donaldson invariants [\wimoore], but a first-principle derivation from the microscopic theory in the general 
case is still lacking -- see however [\sdual], where some steps in this direction have already been taken. 

In eq. \integral\ $\hat T(z)$ is the monodromy-invariant combination:  
$$
{\hat T}(z)= T(z) + { (dz/da)^2 \over 4 \pi {\rm Im} \, \tau },
\eqn\karembeu
$$
where the {\sl contact term} $T(z)$ is given by the general formula 
[\highrank,\strings]: 
$$
T(z)= {4 \over \pi i} {\partial ^2 {\cal F} \over \partial \tau_0^2}.
\eqn\contacto
$$

Here ${\cal F}$ is the prepotential governing the low-energy dynamics of 
the theory. For the asymptotically free theories, $\tau_0$ is defined in terms of the dynamically generated scale $\Lambda_{N_f}$ of the theory by [\strings]: 
$(\Lambda_{N_f})^{4-N_f}=\ex^{i\pi\tau_0}$, while for the finite theories $N_f=4$ and $\cn=4$ it corresponds to the microscopic coupling. For the 
$\cn=4$ theory one gets from \contacto\ 
\REF\zamora{M. Mari\~no and F. Zamora, ``Duality Symmetry 
in Softly Broken $\cn=2$ Gauge Theories", hep-th/9804038.}[\strings] 
-- see also [\zamora] for further details and extensions:  

$$
T(z)= -{1 \over 12} E_2 (\tau) \left( { dz \over da } \right)^2 +
E_2 (\tau_0) {z \over 6} + {m^2 \over 72} E_4 (\tau_0),
\eqn\contato
$$
where $E_2$ and $E_4$ are the Eisenstein functions of weight $2$ and 
$4$, respectively -- see the appendix for further details. 
 
Under a monodromy transformation acting on $\tau$ (holding $\tau_0$ fixed),
 $\tau\to (a\tau+b)/(c\tau+d)$,  
the contact term \contacto\ transforms into itself plus a shift:
$T(z)\to T(z)+ {i\over 2\pi}{c\over c\tau+d}(dz/da)^2$. Under a microscopic duality transformation $\tau_0\to (a\tau_0+b)/(c\tau_0+d)$, the situation is slightly more involved. As these duality transformations interchange the singularities, they induce a non-trivial monodromy transformation 
$\tau\to ({\hat a}\tau+{\hat b})/({\hat c}\tau+{\hat d})$ on the effective low-energy theory [\ferrari]. Under these combined duality transformations 
one has, for example, 
$z\to (c\tau_0+d)^2z$, $(dz/da)\to {(c\tau_0+d)^2\over{\hat c}\tau+
{\hat d} }(dz/da)$, so that [\zamora]:   
$$
T(z)\to (c\tau_0+d)^4 \left(T(z) + {i\over 2\pi}{{\hat c}\over 
{\hat c}\tau+{\hat d}}(dz/da)^2\right) - {i\over\pi}(c\tau_0+d)^3cz
\eqn\hideous
$$

The factor $\Psi$ in \integral\ is essentially the photon partition function, but it contains, apart from the sum over the Abelian line 
bundles of the effective low-energy theory, certain additional terms which carry information about the $2$-observable insertions. In the electric frame 
it takes the form:
$$
\eqalign{
\Psi= \exp \biggl (-{1 \over 4 \pi y} \left({dz \over da}\right)^2 
S_-^2 \biggr)\sum_{\lambda \in \Gamma} 
\,&\biggl[ \lambda\cdot\omega + {i \over 4 \pi y} {dz \over da}
 S\cdot\omega \biggr]
\cr
&
\exp \biggl[ -2i \pi {\overline \tau} (\lambda_+)^2 - 2i\pi\tau (\lambda_-)^2
-2i {dz \over da} S\cdot\lambda_- \biggr], \cr
}
\eqn\lattice
$$
where the lattice $\Gamma$ is $H^2(X,\ZZ)$ shifted by a half-integral class 
$\half\xi=\half w_2(E)$ representing a 't Hooft flux for the $SO(3)$ theory, that is, $\lambda\in H^2(X,\ZZ)+\half w_2(E)$. 
As explained in detail in [\sdual], this shift takes into account the fact 
that in the $SO(3)$ theory, while the rank-$3$ $SO(3)$ bundle $E$ (which 
at a generic vacuum is broken down to $E=(L\oplus L^{-1})^{\otimes 2}=L^2\oplus{\cal O}\oplus L^{-2}$, ${\cal O}$ being a trivial bundle) is always globally defined -- and therefore $L^2$ is represented by an integral class $c_1(L^2)=2\lambda\in 2H^2(X,\ZZ)+w_2(E)$ --, it is not necessarily 
true that the corresponding $SU(2)$ bundle $F$ (which we can somewhat 
loosely represent at 
low energies by $F=L\oplus L^{-1}$) exists, the obstruction being precisely $w_2(E)$: the line bundle $L$ is represented by a class $c_1(L)=\lambda\in H^2(X,\ZZ)+\half w_2(E)$, which is not integral unless $w_2(E)=0~(\mod~2)$. 
If $w_2(E)=0~\mod~2$, the $SO(3)$ bundle can be lifted to an $SU(2)$ bundle 
and one has $F\otimes F = E\oplus{\cal O}$, where now $F$ is a globally 
defined rank-$2$ $SU(2)$ bundle. 

For a given metric in $X$, $\omega$ in \lattice\ is the 
unique -- up to sign -- self-dual two-form satisfying -- see for example [\wimoore,\strings]: $\omega\cdot\omega =1$ (recall that, as explained in [\wimoore], the integral vanishes unless $b^+_2=1$ due to fermion zero modes {\sl and} topological invariance). Here $\cdot$ denotes the intersection pairing on $X$, $\omega\cdot\lambda = \int_X \omega\wedge\lambda$ . Thanks to its properties,  $\omega$ acts as a projector onto the self-dual and antiself-dual subspaces of the two-dimensional cohomology of $X$: $\lambda_+=(\lambda\cdot\omega)\omega$, $\lambda_-=\lambda-\lambda_+$. 
     
From the above formulas it can be readily checked, along the lines explained 
in detail in [\highrank,\strings], that the integral \integral\ is 
well defined and, in particular, is invariant under the monodromy group of 
the low-energy theory (for example, this can be seen almost immediately for 
the semiclassical 
monodromy, which at large $z\simeq u$ takes $z\to \ex^{2\pi i}z$ and 
$a\to -a$, $a_D\to -a_D$, while leaving $\tau\simeq\tau_0$ unchanged.  

\section{Wall crossing at the singular points}

At each of the three singularities, the corresponding local effective 
coupling diverges: $y_{j} = {\rm Im}\,\  \tau_{j}\rightarrow +\infty$, 
$q_{j} \rightarrow 0$. The first step to analyse the behaviour of the 
integral around the singular points is to make a duality transformation 
(in $\tau$) to rewrite the integrand in terms of the appropriate 
variables: $\tau\to -1/\tau$ near the monopole point, etc. 
Due to the divergence of ${\rm Im}\,\  \tau_{j}$, one finds 
a discontinuity in $Z_u$ when $\lambda\cdot\omega$ changes sign. 
We begin by considering the 
behaviour near the semiclassical singularity at $z_1$. As 
the BPS state responsible for the singularity is electrically charged, it is not necessary to perform a duality transformation in this case: the theory is weakly 
coupled in terms of the original effective coupling $\tau$. Let us consider
the integral \integral . Fix $\lambda$ and define $\ell(q)$ as follows:
$$
\ell(q)=f(m,\chi,\sigma,\tau_0) { d z \over  d \tau}
\left({d a \over  dz } \right)^{1- \half \chi  } \Delta^{\sigma/8}
\ex^{2pz+S^2 T(z)-2i(dz/da)S\cdot\lambda}=\sum_{r}c(r)q^{r}.
\eqn\lara
$$
Pick the $n$-th term in the above expansion. The piece of the integral 
relevant to wall crossing is [\wimoore]: 
$$
\int^{\infty}_{y_{\hbox{\sevenrm min}}}{dy\over y^{1/2}}
\int^{+\half}_{-\half} dx c(n) \ex^{2 \pi i x n - 2 \pi yn}
\ex^{-2\pi i x (\lambda_+^2 + \lambda_-^2)}
\ex^{- 2\pi y (\lambda_+^2 - \lambda_-^2)}
\lambda_{+}. 
\eqn\guolcrosin 
$$ 
The integration over $x\equiv\re\,\tau$ imposes $n=\lambda^2$; the 
remaining $y$ integral can be easily evaluated with the result: 

$$
\int^{\infty}_{0}{dy\over y^{1/2}}c(\lambda^2) 
\ex^{-4\pi y\lambda_+^2}\lambda_+={\vert\lambda_+
\vert\over\lambda_+} {c(\lambda^2)\over2} 
\eqn\masguolcrosin
$$
(we have set $y_{\hbox{\sevenrm min}}=0$, as the discontinuity   
comes from the $y\to\infty$ part of the integral). The result of the integral 
\masguolcrosin\ is discontinuous as $\lambda_{+}=\omega
\cdot\lambda\to 0$:  

$$
Z_u\big\vert_{\lambda_+\rightarrow 0^{+}} - Z_u\big\vert_{\lambda_+
\rightarrow 0^{-}} = c(\lambda^2) = \bigl[ q^{-\lambda^2} 
\ell(q) \bigr]_{q^0}=
{\hbox{\rm Res}}_{\,q=0}\bigl[ q^{-\lambda^2-1}\ell(q) \bigr].
\eqn\reguolcrosin
$$

Therefore, the wall-crossing discontinuity of $Z_u$ at 
$z_1$ is:
$$
\eqalign{
&\Delta Z_u\big\vert_{z=z_1}= f(m,\chi,\sigma,\tau_0)\left[ q^{-\lambda^2} 
{d z \over  d \tau}\left({d a \over  dz } \right)^{1-\half\chi} \Delta^{\sigma/8}\ex^{2pz+S^2 T(z)-2i(dz/da)S\cdot\lambda} 
\right]_{q^0}\cr 
&= {\hbox{\rm Res}}_{\,q=0}
f(m,\chi,\sigma,\tau_0)\left[ {dq\over q} q^{-\lambda^2} 
{d z \over  d \tau}\left({d a \over  dz } \right)^{1-\half\chi} \Delta^{\sigma/8}\ex^{2pz+S^2 T(z)-2i(dz/da)S\cdot\lambda} 
\right]\cr 
}
\eqn\finguolcrosin
$$

We have now to evaluate the wall-crossing discontinuities at 
the other two singularities. At the monopole point ($z=z_2$), 
we have to perform a duality transformation to express the 
integral in terms of $\tau_{D}=-1/\tau$, which is the 
appropriate variable there. This duality transformation involves 
a Poisson resummation in \lattice , which exchanges the electric 
class $\lambda\in H^{2}(X;\ZZ)+\half w_{2}(E)$ with the magnetic 
class $\lambda^{*}\in H^{2}(X;\ZZ)$\foot{Notice that, as the manifold 
$X$ is spin, the magnetic class is an integral class, not a 
Spin$_c$ structure as in [\monop].}, and inverts the coupling 
constant $\tau$. The details are not terribly important, so we 
just give the final result for the integral:
$$
\eqalign{
&Z_u = f(m,\chi,\sigma,\tau_0)2^{-b_2/2}
\int{dx_{D}dy_{D}\over{y_D}^{1/2}}{ d z\over  d \tau_{D}}
\bigl({d a_{D} \over  dz } \bigr)^{1- \half \chi  } 
\Delta^{\sigma/8}\ex^{2 p z + S^2 \hat T_{D}-{1\over4\pi y_{D}} 
\left({dz \over da_{D}}\right)^2 S_-^2}\cr &\sum_{\lambda^{*}} 
\left[ {\lambda^{*}\cdot\omega\over2} + {i \over 4 \pi y_{D}} 
{dz \over da_{D}} S\cdot\omega \right](-1)^{\lambda^{*}\cdot\xi}
\ex^{-\half i\pi{\overline
\tau_{D}} (\lambda^{*}_+)^2 - \half i\pi\tau_{D}
(\lambda^{*}_-)^2-i {dz\over da_{D}} S\cdot\lambda^{*}_-},\cr}
\eqn\dualintegral
$$
where now   
$$
{\hat T}_{D}(z)=-{1 \over 12} E_2 (\tau_{D}) \left( { dz \over da_{D} } \right)^2 + E_2 (\tau_0) {z \over 6} + {m^2 \over 72} E_4 (\tau_0)+
 { (dz/da_{D})^2 \over 4 \pi {\rm Im} \, \tau_{D}}.
\eqn\seedorf
$$
The functions $\Delta$ and $z$ are exactly the same as before, but expressed 
in terms of $\tau_D$. The crucial point here is that the modular weight 
of the lattice sum cancels against that of the measure.  

From \dualintegral\ we can easily derive the wall-crossing discontinuity 
at $z_2$ along the lines explained above -- see eqs. 
\lara--\finguolcrosin. The final result differs from \finguolcrosin\ 
in several extra numerical factors:
$$
\eqalign{
\Delta Z_u\big\vert_{z=z_2}&= f(m,\chi,\sigma,\tau_0) 2^{-{b_2\over2}}
(-1)^{\lambda^{*}\cdot\xi}\cr 
{\hbox{\rm Res}}_{\,q_{D}=0}&
\left[ {dq_{D}\over q_{D}}q_{D}^{-{(\lambda^{*})^2\over4}} 
{d z \over  d \tau_{D}}\left({d a_{D}\over  dz } \right)^{1-{\chi\over2}} \Delta^{\sigma/8}\ex^{2pz+S^2 T_{D}-i{dz\over da_{D}}S\cdot\lambda^{*}} 
\right].\cr}
\eqn\finlandia
$$

The corresponding expression at the dyon point $z_3$, is exactly 
the same as \finlandia\ (with $q_{d}$ instead of $q_{D}$) but with 
an extra relative phase $i^{-\xi^2}$ 
%\REF\wijmp{E. Witten, ``Supersymmetric Yang-Mills Theory on a Four-Manifold"
%\journal\jmp&35 (94)5101; hep-th/9403195.}
[\monop,\wimoore,\wijmp], which 
follows from doing the duality transformation 
$\tau\to\tau_{d}=-1/(\tau-1)$ in the lattice sum \lattice .

      €%%%%%%%%%%%%%%%%%%%%%%%%%%%%%%%%%%%%%%%%%%%%%%%%%%%%%%%%%%

       %%%%%%%%%%%%%%%%%%%%%%%%%%%%%%%%%%%%%%%%%%%%%%%%%%%%%%%%%%

\section{Contributions from the singularities}

At each of the singularities, the complete effective theory 
contains a dual Abelian vector multiplet\foot{This is so for the monopole and dyon singularities; at the semiclassical singularity, the distinguished 
vector multiplet is the original electric one.} (weakly) coupled to a 
massless charged hypermultiplet representing the BPS configuration 
responsible for the singularity. This effective theory can be twisted in the 
standard way, and the resulting topological theory is the celebrated 
Witten's Abelian monopole theory. Its moduli space is defined by the Abelianized version of eqs. \masecuaciones . On spin four-manifolds, 
and for a given gauge configuration $\tilde\lambda\in H^{2}(X;\ZZ)$, 
the virtual dimension of the moduli space can be seen to be
$$ 
{\hbox{\rm dim}}_{\tilde\lambda}= -{(2\chi+3\sigma)\over4}+
(\tilde\lambda)^{2}.
\eqn\indiecita
$$
A class $\tilde\lambda$ for which ${\hbox{\rm dim}}_{\tilde\lambda}=0$ 
is called a {\sl basic} class. If we define $x=-2\tilde\lambda$, we see 
from \indiecita\ that for a basic class $x\cdot x=2\chi+3\sigma$. As ${\hbox{\rm dim}}_{x}=0$, 
the moduli space consists (generically) of a (finite) collection of 
isolated points. The partition function of the theory 
evaluated at each basic class gives the Seiberg-Witten invariant 
$n_{x}$. The complete partition function will therefore be a (finite) sum 
over the different basic classes: $Z_{{\hbox{\sevenrm singularity}}}\sim 
\sum_{x}n_{x}$. If, on the other hand, the dimension of the moduli space 
of Abelian monopoles is strictly positive, one has to insert 
observables to obtain a non-trivial result. This leads to the definition  
of the generalized Seiberg-Witten invariants 
\REF\taubes{C.H. Taubes, ``The Seiberg-Witten Invariants and the 
Gromov Invariants", {\sl Math. Res. Lett.} {\bf 2} (1995), 221.}[\wimoore,\taubes]: if 
${\hbox{\rm dim}}_{\tilde\lambda}= 2n$ (otherwise the invariant is automatically set to zero), 
$$
SW_n(\tilde\lambda)=\left\langle(\tilde\phi)^n\right\rangle_
{\tilde\lambda},
\eqn\general
$$
where $\tilde\phi$ is the (twisted) scalar field in the Abelian $\cn=2$ 
vector multiplet. For a four-manifold $X$ with $b^{+}_{2}>1$, the $u$-plane integral vanishes and the only contributions to the topological correlation functions come from the effective theories at the singularities. Those  manifolds with $b^{+}_{2}>1$ for which the only non-trivial contributions 
come from the zero-dimensional Abelian monopole moduli spaces are called of 
{\sl simple type}. No four-manifold with $b^{+}_{2}>1$ is known which is not of simple type. We will restrict ourselves to manifolds of simple type. The generalization to positive-dimensional monopole moduli spaces should be straightforward from the explicit formulas in [\wimoore] and our own results. 
 
The general form of the contribution to the generating function 
$\left\langle\ex^{p{\cal O}+I(S)}\right\rangle_{\xi}$ from  
the twisted Abelian monopole theory at a given singularity was presented in [\wimoore]. It contains certain effective gravitational couplings as well as 
contact terms among the observables. We just adapt here eq. (7.12) of 
[\wimoore]:   

$$
\eqalign{
&\left\langle \ex^{p{\cal O}+I(S)}\right\rangle_{\tilde\lambda_j, z_j,\xi} 
= SW_n(\tilde\lambda)\, {\hbox{\rm Res}}_{\,a_j=0}\Bigg\{{da_{j}\over (a_{j})^{1+\tilde
\lambda_{j}^2/2-(2\chi+3\sigma)/8}}(-1)^{\tilde\lambda_{j}\cdot\xi}\cr 
&\ex^{2pz - i {d z \over da_{j}} \tilde\lambda_{j}\cdot S 
+ S^2 T_{j}(z)} C_j(z)^{-\tilde\lambda_j^2} P_j(z)^{\sigma/8}
L_j(z)^{\chi/4}\Bigg\}.
\cr}
\eqn\sw
$$
In \sw, $a_{j}$ is the distinguished (dual) coordinate at the singularity:
$a-a(z_1)\simeq a-m/\raiz$ at the semiclassical singularity, 
$a_{D}-a_{D}(z_2)$ at the monopole point, and 
$(a-a_{D})-(a-a_{D})(z_3)$ at the dyon point. $C_j$, $P_j$, $L_j$ are a priori unknown functions, which can be determined by wall crossing as follows [\wimoore]. For $b^{+}_2=1$ and  fixed $\tilde\lambda_{j}$, \sw\ exhibits a wall-crossing behaviour when $\omega\cdot\tilde\lambda_j$ changes sign. 
At such points, 
the only discontinuity comes from $SW_n(\tilde\lambda)$, which jumps 
by $\pm 1$. 
Therefore, the discontinuity in \sw\ is: 
$$
\eqalign{
&\Delta\left\langle \ex^{p{\cal O}+I(S)}\right\rangle_{\tilde\lambda_j, z_j,\xi} = \pm {\hbox{\rm Res}}_{\,a_j=0}\Bigg\{{da_{j}\over (a_{j})^
{\tilde\lambda_{j}^2/2-\sigma/8}}(-1)^{\tilde\lambda_{j}\cdot\xi}
\cr &\ex^{2pz - i {d z \over da_{j}} \tilde\lambda_{j}\cdot S 
+ S^2 T_{j}(z)} C_j(z)^{-\tilde\lambda_j^2} P_j(z)^{\sigma/8}
L_j(z)^{1-\sigma/4}\Bigg\}.
\cr}
\eqn\swdiscont
$$
(We have set $\chi+\sigma=4$, which is equivalent to $b^{+}_{2}=1$ 
for $b_{0}(X)=1, b_{1}(X)=0$.) The crucial point now is that the 
complete expression for the generating function cannot have 
wall-crossing discontinuities at finite values of $z$. This is not 
difficult to understand if one realizes that nothing physically (or 
mathematically) special occurs at the singular points: when expressed 
in terms of the appropriate variables, and once all the relevant degrees 
of freedom are taken into account, the low-energy effective description 
is perfectly smooth there. The conclusion is therefore that the 
discontinuity in the $u$-plane integral has to cancel against the 
discontinuity in the contribution from the effective theory at the 
singularity. As shown in [\wimoore], this suffices to fix the unknown 
functions $C_j$, $P_j$, $L_j$ in \sw .

At a generic vacuum, the $SU(2)$ -- or, more generally, the $SO(3)$ --  
rank-$3$  bundle $E$ is broken down to $E=L^2\otimes{\cal O}
\otimes L^{-2}$ by the 
Higgs mechanism, where ${\cal O}$ is the trivial line bundle 
(where the neutral $\cn=4$ multiplet lives), 
while $L^{\pm 2}$ are globally defined line bundles where the charged 
massive $W^{\pm}$ $\cn=4$ multiplets live. With our conventions, $c_1(L^2)=2c_1(L)=2\lambda\in 2H^{2}(X;\ZZ)+w_2(E)$, which is indeed an integral class. The ``monopole" becoming massless at the semiclassical singularity is just one of the original electrically charged (massive) 
quarks, which sits in an $\cn=4$ Abelian multiplet together with the $\cn=2$ vector multiplet of one of the massive $W$ bosons. The corresponding basic classes are therefore of the form: 
$$
x=-2\tilde\lambda_{1}=-2c_1(L^2)=-4\lambda\in 4H^{2}(X;\ZZ)+2w_2(E),
\eqn\alice
$$
which are even classes since the manifold $X$ is spin\foot{If the manifold is not spin, the basic classes are shifted from being even classes by the second Stieffel-Whitney class of the manifold, $w_2(X)$ [\monop].}. 
Notice that, because 
of \alice , not all the basic classes of $X$ will contribute to the 
computation at $z_1$. Rather, only those basic classes $x$ satisfying 
$$
{x\over2}+w_2(E)= 0~(\mod~2)\Leftrightarrow \left[{x\over2}\right]=w_2(E), 
\eqn\constraint
$$
with $\left[{x\over2}\right]$ the mod $2$ reduction of ${x\over2}$, can give 
a non-zero contribution.  

Taking this into account, we can rewrite \swdiscont\ at $z_1$ as follows:
$$
\eqalign{
&\Delta\left\langle \ex^{p{\cal O}+I(S)}\right\rangle_{2\lambda,z_1,\xi} 
= \pm {1\over\pi i}{\hbox{\rm Res}}_{\,q=0}\Bigg\{{dq \over q} (a_{1})^
{-2\lambda^2+\sigma/8}{da\over dz}{dz\over d\tau}
\cr &\ex^{2pz - 2i {d z \over da} \lambda\cdot S 
+ S^2 T_{1}(z)} C_1(z)^{-4\lambda^2} P_1(z)^{\sigma/8}
L_1(z)^{1-\sigma/4}\Bigg\} 
\cr}
\eqn\swdiscuno
$$
(notice that the phase $(-1)^{\tilde\lambda\cdot\xi}$ does not appear 
here), where we have used ${\hbox{\rm Res}}_{\,a=0}\left[da F(a)\right]=
2\,{\hbox{\rm Res}}_{\,q=0}\left[dq (da/dq) F(a)\right]$, and we have 
taken into account that, near $z=z_1$, $a_{1}=a-a(z_1)=a_0 q^{1/2}+\cdots$.
By comparing \swdiscuno\ with the wall-crossing formula for the integral 
at $z_1$, \finguolcrosin , we can determine the unknown functions in \sw . 
We find, for example, 
$$
\eqalign{
T_1& = T,\cr
(C_1)^4 & = {a_1{}^2\over q} \cr},
\qquad\qquad
\eqalign{
L_1 & =\left({dz \over  da}\right)^2,\cr
P_1 & = {\Delta\over a_1}. \cr}
\eqn\alessia
$$

Putting everything together, the final form for the contribution to 
the generating function at $z_1$ is given by the following formula:
$$
\eqalign{
\left\langle \ex^{p{\cal O}+I(S)}\right\rangle_{\lambda,z_1,\xi}&= 
SW_n(\tilde\lambda)\, 2\pi if(m,\chi,\sigma,\tau_0)\cr 
{\hbox{\rm Res}}_{\,q=0}&
\left[ dq q^{-\lambda^2} 
{d z \over  d q}\left({d a \over  dz } \right)^{1-\half\chi} a_{1}{}^{{\chi+\sigma\over4}-1}\Delta^{\sigma/8}\ex^{2pz+S^2 T(z)-2i(dz/da)S\cdot\lambda} 
\right].\cr 
}
\eqn\seibergwitten
$$

We can now specialize to the simple-type case, for which $4\lambda^2 = 
(2\chi+3\sigma)/4$. We use the following series expansions around $z_1$:
$$
\eqalign{
z&=z_1+\kappa_1 q^{\half}+\cdots,\cr 
a_{1}&= (da/dz)_1(z-z_1)+\cdots = 
(da/dz)_1 \kappa_1 q^{\half}+\cdots,\cr
da/dz &=(da/dz)_1+\cdots,\cr
\Delta &={\eta(\tau)^{12}\over 2^3 (da/dz)^6}= 
2^{-3} (dz/da)_1{}^6 q^{\half}+\cdots.\cr}
\eqn\serexp
$$
The final formula is the following:
$$
\eqalign{
\left\langle \ex^{p{\cal O}+I(S)}\right\rangle_{x,z_1,\xi}&= 
2^{-{3\sigma\over8}}\pi i
f(m,\chi,\sigma,\tau_0)\, \cr (\kappa_1)^{\nu} 
&\left({d a \over  dz } \right)_1^{-(\nu+\sigma/4)} 
\ex^{2pz_1+S^2 T(z_1)}\,
\delta_{\left[{x\over2}\right],\xi}\,n_{x}\,
\ex^{\half i(dz/da)_1S\cdot x},  
\cr 
}
\eqn\moorewitten
$$
where $\nu=(\chi+\sigma)/4$. The delta function 
$\delta_{\left[{x\over2}\right],\xi}$ in \moorewitten\ enforces the 
constraint \constraint, and $T(z_1)$ is given by:
$$
T(z_1)= -{1 \over 12} ( dz/da)_1{}^2 +
E_2 (\tau_0) {z_1 \over 6} + {m^2 \over 72} E_4 (\tau_0).
\eqn\contralto
$$

The corresponding expressions at the monopole and dyon 
singularities can be determined along the same lines. One 
has to take into account the relative factors in each case, 
and the fact that, at these singularities, the basic classes 
are given by $x=-2\tilde\lambda=-2\lambda^{*}$, where 
$\lambda^{*}$ is the appropriate dual class. One finds in this 
way, for the monopole singularity at $z_2$, the following expression: 
$$
\eqalign{
\left\langle \ex^{p{\cal O}+I(S)}\right\rangle_{x,z_2,\xi}&= 
2^{-{3\sigma\over8}-{b_2\over 2}}\pi i
f(m,\chi,\sigma,\tau_0)\, \cr (-1)^{\sigma/8}(\kappa_2)^{\nu} 
&\left({d a_{D} \over  dz } \right)_2^{-(\nu+\sigma/4)} 
\ex^{2pz_2+S^2 T(z_2)}
(-1)^{ {x\over2}\cdot \xi}\,n_{x}\,\ex^{\half i(dz/da_{D})_2 S\cdot x},  
\cr 
}
\eqn\monopolo
$$
while for the dyon singularity at $z_3$ one finds:
$$
\eqalign{
\left\langle \ex^{p{\cal O}+I(S)}\right\rangle_{x,z_3,\xi}&= 
2^{-{3\sigma\over8}-{b_2\over 2}}\pi i
f(m,\chi,\sigma,\tau_0)\, \cr i^{-\xi^2}(\kappa_3)^{\nu} 
&\left({d (a_{D}-a)\over  dz } \right)_3^{-(\nu+\sigma/4)} 
\ex^{2pz_3+S^2 T(z_3)}
(-1)^{{x\over2}\cdot\xi}\, n_{x}\,\ex^{\half i(dz/d(a_{D}-a))_3 S\cdot x},  
\cr 
}
\eqn\dyon
$$
where $T(z_2)$ and $T(z_3)$ are given by expressions analogous to \contralto :
$$
\eqalign{
T(z_2)&= -{1 \over 12} ( dz/da_{D})_2{}^2 +
E_2 (\tau_0) {z_2 \over 6} + {m^2 \over 72} E_4 (\tau_0),\cr
T(z_3)&= -{1 \over 12} ( dz/d(a_{D}-a))_3{}^2 +
E_2 (\tau_0) {z_3 \over 6} + {m^2 \over 72} E_4 (\tau_0). 
\cr}
\eqn\mezzosoprano
$$

%%%%%%%%%%%%%%%%%%%%%%%%%%%%%%%%%%%%%%%%%%%%%%%%%%%%%%%%%%%%%%%

%%%%%%%%%%%%%%%%%%%%%%%%%%%%%%%%%%%%%%%%%%%%%%%%%%%%%%%%%%%%%%%

\section{The formula for the generating function}

The complete formula for the generating function of the half-twisted 
theory on simply-connected spin four-manifolds of simple type is given 
by the combination of \moorewitten , \monopolo\ and \dyon , summed over 
the basic classes (we do not sum over 't Hooft fluxes, though). The 
contribution from the $u$-plane integral is absent, since it vanishes for manifolds with $b^{+}_2>1$. As for the as yet unknown function 
$f(m,\chi,\sigma,\tau_0)$, it is not possible to determine it completely
in the context of the $u$-plane approach. 
However, we will propose an ansatz for this function, which is motivated by a series of natural conditions that it has to satisfy. We will discuss later how modifications of the proposed form for $f(m,\chi,\sigma,\tau_0)$ violate those conditions.
As we will show, our ansatz leads to the right mass dependence according to our previous arguments, which led to \coop, and makes the partition function display  two properties of the partition function of the twisted $\cn=4$ supersymmetric theory
considered by Vafa and Witten [\vafa]: it is a modular form of weight 
$-\chi/2$ and contains the Donaldson invariants in the form shown 
in [\coreatres]. In addition, its final expression reduces to the Vafa-Witten partition function on $K3$. 

Our ansatz for $f(m,\chi,\sigma,\tau_0)$, which turns out to satisfy the properties stated above, is:
$$
f(m,\chi,\sigma,\tau_0)= -{i\over\pi}\,2^{(3\chi+7\sigma)/8}\, m^{\sigma/8} 
\eta(\tau_0)^{-12\nu}.
\eqn\ansatz
$$ 
Taking \moorewitten , \monopolo\ and \dyon, the formula that one obtains for the generating function of all the topological correlation functions for simply-connected spin manifolds is the following:
$$
\eqalign{
\left\langle\ex^{p{\cal O}+I(S)}\right\rangle_{\xi}= 2^{\nu/2}
2^{(2\chi+3\sigma)/8} m^{\sigma/8} (\eta(\tau_0))^{-12\nu}&\Bigg\{
\cr 
(\kappa_1)^{\nu} \left ({{da}\over{dz}}\right)^
{-(\nu+{\sigma\over4})}_1\ex^{2pz_1 + S^2 T_1} &\sum_{x} 
\delta_{\left[{x\over2}\right],\xi} \,n_x \,\ex^{{i\over2} \left (dz
/da\right)_1 x\cdot S}\cr +
2^{-{b_2\over2}}(-1)^{\sigma/8}(\kappa_2)^{\nu} \left 
({{da_D}\over{dz}}\right)^
{-(\nu+{\sigma\over4})}_2\,\ex^{2pz_2 + S^2 T_2} &\sum_{x} 
(-1)^{\xi \cdot {x\over2}}\,n_x\, \ex^{{i\over2} \left (dz/
da_D\right)_2 x\cdot S}\cr +
2^{-{b_2\over2}} i^{-\xi^2}(\kappa_3)^{\nu} \left 
({{d(a_D-a)}\over{dz}}\right)^
{-(\nu+{\sigma\over4})}_3\ex^{2pz_3 + S^2 T_3} &\sum_{x} 
(-1)^{\xi \cdot {x\over2}}\,n_x\, \ex^{{i\over2} \left (dz/
d(a_D-a)\right)_3 x\cdot S}\Bigg\},\cr} 
\eqn\formulauno
$$
where the sum $\sum_{x}$ is over {\sl all} the Seiberg-Witten basic 
classes. This formula can be written in terms of modular 
forms by substituting the explicit expressions \yespadas\ for 
$\kappa_{j}$ and \zhivago\ for the periods. Notice that there is no need 
to resolve the square roots in \zhivago . Indeed, the periods in 
\formulauno\ are raised to the power $-(\nu+\sigma/4)$. Since the 
manifold $X$ is spin, $\sigma=0~\mod~8$, so $\sigma/4$ is even. As 
for $\nu=(\chi+\sigma)/4$, it is only guaranteed\foot{
For $x=-2\tilde\lambda=-2c_1(\tilde L)$ a basic class, $\nu$ is the index 
of the corresponding Dirac operator\hfill\break 
$D:\Gamma(X,S^{+}\otimes\tilde L)\to 
\Gamma(X,S^{-}\otimes\tilde L)$, which is always an integer [\monop].}
that $\nu\in\ZZ$. 
Nevertheless, as explained in sect. 11.5 of [\wimoore], one can 
take advantage of the following property of the Seiberg-Witten invariants $n_{x}$ -- see [\monop] for a quick proof:
$$
n_{-x}=(-1)^{\nu}n_{x}. 
\eqn\sherezade    
$$
Upon summing over $x$ and $-x$ using \sherezade, the factors $\ex^{{i\over2}(dz/da)_{j}x\cdot S}$ average to a cosine 
when $\nu$ is even, and to 
a sine when $\nu$ is odd. Therefore, no odd powers of $(dz/da)_j$ appear 
in \formulauno. Note that as $x$ is an even class on 
spin manifolds, $\xi \cdot x/2$ is always an integer, and therefore 
the phase $(-1)^{\xi \cdot {x\over2}}$ does not spoil the argument. 
Likewise, if $x+2\xi=0~\mod~4$, it is also true that 
$-x+2\xi=0~\mod~4$ ($x$ is an even class), so if a given basic class 
$x$ contributes to the first term in \formulauno , so does $-x$.

We will now work out a more explicit formula for the partition function
(setting $p=0$ and $S=0$ in \formulauno). Notice that, since the partition function does not care 
about whether the manifold is simply-connected or not, at least in the simple-type case (in any case, we are not computing correlation 
functions of observables), we can easily extend 
our result to the non simply-connected case. As in [\vafa] this extension involves the introduction of a factor $2^{-b_1}$ in two of the three contributions. Notice also that, because 
of \sherezade , the partition function is zero when $\nu$ is odd, 
since $n_{x}+n_{-x}=n_{x}(1+(-1)^{\nu})$. Therefore, in the following 
formula for the partition function we assume that $\nu$ is even. Upon 
substituting eqs. \yespadas\ and \zhivago\ in \formulauno , and taking 
into account the identities  (A10), the partition function 
for a fixed 't Hooft flux $\xi$ is given by:
$$
\eqalign{
Z_\xi &= m^{3(2\chi+3\sigma)/8}\Bigg\{
\left({G(q_0{}^2)\over4}\right)^{\nu/2}
\left(\vartheta_3\vartheta_4\right)^{(2\chi +3\sigma)/2}
\sum_{x}\delta_{\xi,\left[{x\over2}\right]}n_x
\cr
&+ 2^{1-b_1+{1\over4}(7\chi +11\sigma)} 
(-1)^{{\sigma\over8}}\left(G(q_0{}^{1/2})\right)^{\nu/2}
\left({\vartheta_2 \vartheta_3\over2}\right)^
{(2\chi +3\sigma)/2}\sum_{x}(-1)^{{x\over2}\cdot \xi} n_x 
\cr &
+ 2^{1-b_1+{1\over4}(7\chi +11\sigma)} 
(-1)^{{\sigma\over8}}  i^{-\xi^2}
\left(G(-q_0{}^{1/2})\right)^{\nu/2}
\left({\vartheta_2 \vartheta_4\over2}\right)^
{(2\chi +3\sigma)/2}
\sum_{x} (-1)^{{x\over2}\cdot \xi}n_x\Bigg\}.
\cr
}
\eqn\formulados
$$ 

The partition function for gauge groups $SU(2)$ and $SO(3)$ is easily 
obtained from this expression. One finds:
$$
\eqalign{
Z_{SU(2)} &= Z_{\xi=0}/2^{1-b_1} \cr
   &= m^{3(2\chi+3\sigma)/8}\Bigg\{2^{b_1-1}
\left({G(q_0{}^2)\over4}\right)^{\nu/2}
\left(\vartheta_3\vartheta_4\right)^{(2\chi +3\sigma)/2}
\sum_{x=4y}n_x
\cr
&+ 2^{{1\over4}(7\chi +11\sigma)} 
(-1)^{\sigma/8}\left(G(q_0{}^{1/2})\right)^{\nu/2}
\left({\vartheta_2 \vartheta_3\over2}\right)^
{(2\chi +3\sigma)/2}\sum_{x}n_x 
\cr &
+ 2^{{1\over4}(7\chi +11\sigma)} 
(-1)^{\sigma/8} 
\left(G(-q_0{}^{1/2})\right)^{\nu/2}
\left({\vartheta_2 \vartheta_4\over2}\right)^
{(2\chi +3\sigma)/2}
\sum_{x}n_x\Bigg\}.
\cr
}
\eqn\formulacinco
$$ 
The constraint $x=4y$ in the first term in \formulacinco\ means that 
one has to consider only those basic classes $x\in 4H^2(X;\ZZ)$. 
Notice that this constraint implies that the corresponding 
contribution vanishes unless $x^2 = 2\chi+3\sigma = 8\nu +\sigma = 
16 y^2 = 0~\mod~32$. But since $\nu$ is even, this requires 
$\sigma = 0~\mod~16$. Therefore, when $\sigma\in 16\ZZ$, the first 
singularity contributes to the $SU(2)$ partition function, and the 
leading behaviour of the partition function is: 
$$
Z_0\sim q_0{}^{-\nu}+{\cal O}(q_0{}^{-\nu+1}).
\eqn\leading
$$
As in [\vafa], the leading contribution could be interpreted as 
the contribution of the trivial connection, shifted from $(q_0)^{0}$ to 
$q_0{}^{-\nu}$ by the $c$-number we referred to in sect. 3. But notice 
that the next power in the series expansion is $q_0{}^{-\nu+1}$. The gap 
between the trivial connection contribution and the first non-trivial 
instanton contribution which was noted in [\vafa] for the Vafa-Witten 
partition function is not present here: all instanton configurations 
contribute to $Z_{SU(2)}$.  
 
On the other hand, when $\sigma= 8(2k + 1),~k\in\ZZ$, the first 
singularity does not contribute to the partition function and the leading 
behaviour comes from the strong coupling singularities. Then $Z_0$ has an 
expansion:    
$$
Z_0\sim q_0{}^{{2\chi+3\sigma\over16}}+{\cal O}
(q_0{}^{{2\chi+3\sigma\over16}+1}), 
\eqn\masleading
$$
again with no gap between the contribution of the trivial connection 
(shifted from $(q_0)^{0}$ by the $c$-number $q_0{}^{{2\chi+3\sigma\over16}}$) 
and higher-order instanton contributions. 

As for the $SO(3)$ partition function, one has to sum \formulados\ 
over all allowed bundles, which means summing over all allowed 
't Hooft fluxes. One finds in this way:
$$
\eqalign{
Z_{SO(3)} &= \sum_{\xi}Z_{\xi}=\cr
          & m^{3(2\chi+3\sigma)/8}\Bigg\{
\left({G(q_0{}^2)\over4}\right)^{\nu/2}
\left(\vartheta_3\vartheta_4\right)^{(2\chi +3\sigma)/2}
\sum_{x}n_x
\cr
&+ 2^{1-b_1+b_2+{1\over4}(7\chi +11\sigma)} 
(-1)^{\sigma/8}\left(G(q_0{}^{1/2})\right)^{\nu/2}
\left({\vartheta_2 \vartheta_3\over2}\right)^
{(2\chi +3\sigma)/2}\sum_{x=4y}n_x 
\cr &
+ 2^{1-b_1+b_2/2+{1\over4}(7\chi +11\sigma)} 
\left(G(-q_0{}^{1/2})\right)^{\nu/2}
\left({\vartheta_2 \vartheta_4\over2}\right)^
{(2\chi +3\sigma)/2}
\sum_{x} n_x\Bigg\}.
\cr
}
\eqn\formulario
$$ 

To perform the summation over fluxes in \formulario, one uses the 
following identities [\vafa]:
$$
\eqalign{
\sum_{\xi}\sum_{x}n_{x}\,
\delta_{\left[{x\over2}\right],\xi}&=\sum_{x}\,n_{x},\cr
\sum_{\xi}\sum_{x}
(-1)^{{x\over2}\cdot\xi}\,n_{x}&=
2^{b_2}\sum_{x=4y}n_{x},\cr
\sum_{\xi}\,i^{-\xi^2}\sum_{x}
(-1)^{{x\over2}\cdot\xi}\,n_{x}&=2^{b_2/2}
(-1)^{\sigma/8}\sum_{x}\,n_{x}.\cr}
\eqn\profident
$$

\endpage

\chapter{Duality transformations of the generating function}

In this section we will study the properties under duality transformations of the generating function \formulauno. We will start by checking the modular properties of $Z_{\xi}(\tau_0)$ as given in \formulados. 
As explained in [\vafa], one should expect the following behaviour 
under the modular group. Under a $T$ transformation, taking 
$\tau_0\to\tau_0+1$, the partition function for fixed $\xi$ must transform 
into itself with a possible anomalous $\xi$-dependent phase. Indeed, 
\formulados\ behaves under $T$ in the expected fashion:
$$
Z_{\xi}(\tau_0+1)=i^{-\xi^2}(-1)^{\sigma/8}Z_{\xi}(\tau_0).  
\eqn\tduality
$$
Checking \tduality\ involves some tricky steps that we now explain. 
Let us rewrite \formulados\ as:   
$$
Z_\xi =A_1(\tau_0)\sum_{x}\delta_{\xi,\left[{x\over2}\right]}n_x + 
\left [ A_2(\tau_0)+
i^{-\xi^2}A_3(\tau_0)\right]\sum_{x} (-1)^{{x\over2}\cdot \xi}n_x.
\eqn\fmdos
$$ 
where we have put $A_1(\tau_0)=m^{3(2\chi+3\sigma)/8}\left(G(q_0{}^2)/4\right)^{\nu/2}
\left(\vartheta_3\vartheta_4\right)^{(2\chi +3\sigma)/2}$, 
$A_2(\tau_0)=m^{3(2\chi+3\sigma)/8}2^{1-b_1+{1\over4}(7\chi +11\sigma)} 
(-1)^{\sigma/8}\left(G(q_0{}^{1/2})\right)^{\nu/2}
\left(\vartheta_2 \vartheta_3/2\right)^
{(2\chi +3\sigma)/2}$, and so on. Under $\tau_0\to\tau_0+1$ we have: 
$$
A_1\to A_1,\quad A_2\to \ex^{{i\pi\over8}(2\chi+3\sigma)}A_3=
(-1)^{{\sigma\over8}}A_3,\quad \quad A_3\to \ex^{{i\pi\over8}(2\chi+3\sigma)}A_2=
(-1)^{{\sigma\over8}}A_2, 
\eqn\procyon
$$
and we have taken into account that $\nu\in 2\ZZ$ throughout. In view of 
\procyon, \fmdos\ transforms as follows:
$$
Z_\xi\too \tilde Z_\xi= A_1(\tau_0)\sum_{x}\delta_{\xi,\left[{x\over2}\right]}n_x + 
(-1)^{\sigma/8}\left [ A_3(\tau_0)+
i^{-\xi^2}A_2(\tau_0)\right]\sum_{x} (-1)^{{x\over2}\cdot \xi}n_x, 
\eqn\alcyone
$$ 
that is 
$$
\tilde Z_\xi= i^{-\xi^2}(-1)^{{\sigma\over8}}\left(i^{\xi^2}(-1)^{{\sigma\over8}}  A_1\sum_{x}\delta_{\xi,\left[{x\over2}\right]}n_x + 
\left [ A_2+
i^{-\xi^2}A_3\right]\sum_{x} (-1)^{{x\over2}\cdot \xi}n_x\right). 
\eqn\deneb
$$ 
The phase $i^{\xi^2}(-1)^{\sigma/8}$ in front of $A_1$ seems to spoil 
the invariance of $Z_\xi$ under $T$. That this is not actually so is 
because of the constraint $\xi+x/2=0~\mod~2$. Indeed, 
when $\xi+x/2=0~\mod~2$ we have $\xi^2=x^2/4 ~\mod~4$, and therefore 
$$
i^{\xi^2}=i^{x^2/4}=i^{(2\chi+3\sigma)/4}=i^{2\nu+\sigma/4}=(-1)^{\sigma/8}, 
\eqn\vega
$$
which guarantees that $Z_\xi$ is invariant (up to a phase) under $\tau_0
\to\tau_0+1$.  

Likewise, under an $S$ transformation, taking $\tau_0\to -1/\tau_0$,  
one should expect the following behaviour:
$$
Z_{\xi}(-1/\tau_0)\propto\sum_{\zeta} (-1)^{\zeta\cdot\xi}
Z_{\zeta}(\tau_0).  
\eqn\thooftrule
$$
It turns out that the partition function \formulados\ indeed  
satisfies \thooftrule . In fact, 
$$
Z_{\xi}(-1/\tau_0) = 2^{-b_2/2}(-1)^{\sigma/8}
\left({\tau_0\over i}\right)^{-\chi/2}
\sum_{\zeta}(-1)^{\zeta\cdot\xi}Z_{\zeta}(\tau_0).
\eqn\thooft
$$
In order to check \thooft , one has to use the modular properties 
of the different functions in \formulados, which are compiled in the 
appendix, and handle with care the summation over 
fluxes in \thooft . 
The following identities are useful:
$$
\eqalign{
\sum_{\zeta}(-1)^{\zeta\cdot\xi}\sum_{x}n_{x}\,
\delta_{\left[{x\over2}\right],\zeta}&=\sum_{x}
(-1)^{{x\over2}\cdot\xi}\,n_{x},\cr
\sum_{\zeta}(-1)^{\zeta\cdot\xi}\sum_{x}
(-1)^{{x\over2}\cdot\zeta}\,n_{x}&=
2^{b_2}\sum_{x}n_{x}\,\delta_{\left[{x\over2}\right],\xi},\cr
\sum_{\zeta}(-1)^{\zeta\cdot\xi}\,i^{-\zeta^2}\sum_{x}
(-1)^{{x\over2}\cdot\zeta}\,n_{x}&=2^{b_2/2}i^{-\xi^2}
(-1)^{\sigma/8}\sum_{x}(-1)^{{x\over2}\cdot\xi}
\,n_{x}.\cr}
\eqn\profiden
$$
To prove these identities, we borrow from eq. (5.40) of   
[\vafa] the following formulas:  
$$
\eqalign{
&\sum_{z}(-1)^{z\cdot y} \delta_{z,z'}
=(-1)^{y\cdot z'},\cr
&\sum_{z}(-1)^{z\cdot y}
= 2^{b_2}\delta_{y,0},\cr
&\sum_{z}(-1)^{z\cdot y} i^{-z^2}
= 2^{b_2/2}i^{-\sigma/2 + y^2}.
\cr}
\eqn\sumident
$$
In addition to this, we have to take into account that, since the 
manifold is spin, 
$\sigma=0~\mod~8$, $x\in 2H^{2}(X;\ZZ)$, and $x\cdot\xi,\xi^2\in 2\ZZ$. 
Similarly, one should not forget to impose the constraint 
$\nu\in 2\ZZ$. 

Using \formulacinco\ and \formulario, one finds the following duality transformation properties for the $SU(2)$ and $SO(3)$ partition 
functions:

$$
\eqalign{
Z_{SU(2)}(\tau_0+1)&=(-1)^{\sigma/8}Z_{SU(2)}(\tau_0),\cr\cr
Z_{SO(3)}(\tau_0+2)&= Z_{SO(3)}(\tau_0),\cr\cr
Z_{SU(2)}(-1/\tau_0)&=(-1)^{\sigma/8}2^{-\chi/2}\tau_{0}{}^{-\chi/2}
Z_{SO(3)}(\tau_0).\cr}
\eqn\tdual
$$
As expected, the partition function for $SO(3)$ does not transform properly under $\tau_0\rightarrow \tau_0+1$, but transforms into itself under $\tau_0\to\tau_0+2$. The last of these 
three equations corresponds precisely to the strong-weak coupling 
duality transformation conjectured by Montonen and Olive [\monoli].

We will now analyse the behaviour of the topological correlation functions 
under modular transformations. First, it is useful to work out how the different terms entering \formulauno\ transform. Under $\tau_0\to-1/\tau_0$ we have, 
$$
\eqalign{
z_1&\too \tau_0{}^2 z_2,\cr
z_2&\too \tau_0{}^2 z_1,\cr
z_3&\too \tau_0{}^2 z_3,\cr
}\qquad\qquad
\eqalign{
T_1&\too \left({\tau_0\over i}\right)^4\left( 
T_2-{i\over\pi\tau_0}z_2\right),\cr 
T_2&\too \left({\tau_0\over i}\right)^4\left( 
T_1-{i\over\pi\tau_0}z_1\right),\cr 
T_3&\too \left({\tau_0\over i}\right)^4\left( 
T_3-{i\over\pi\tau_0}z_3\right).
\cr}
\eqn\testadicazzo
$$ 
These formulas entail for the topological correlation functions the 
following behaviour under an $S$ transformation:
$$
\eqalign{
\left\langle{1\over8\pi^2}\tr\,\phi^2\right\rangle
^{SU(2)}_{\tau_0}&=\left\langle {\cal O}
\right\rangle^{SU(2)}_{\tau_0}={1\over\tau_0{}^2}\left\langle 
{\cal O}\right\rangle^{SO(3)}_{-1/\tau_0},\cr\cr
\left\langle{1\over8\pi^2}\int_S\tr\,\left(2\phi F + \psi\wedge\psi\right)\right\rangle^{SU(2)}_{\tau_0}&=
\left\langle I(S)\right\rangle^{SU(2)}_{\tau_0}= 
{1\over\tau_0{}^2}\left\langle I(S)\right\rangle^{SO(3)}_{-1/\tau_0},
\cr\cr
\left\langle I(S)I(S)\right\rangle^{SU(2)}_{\tau_0}= 
\left({\tau_0\over i}\right)^{-4}&\left\langle I(S)I(S)\right\rangle^{SO(3)}_{-1/\tau_0}+{i\over2\pi}{1\over\tau_0{}^3}
\left\langle {\cal O}\right\rangle^{SO(3)}_{-1/\tau_0}\sharp 
(S\cap S).\cr
}
\eqn\bizarre
$$

At first sight, the behaviour of $\left\langle I(S)\right\rangle$ under $\tau_0\to-1/\tau_0$ seems rather unnatural. Since $I(S)$ is essentially 
the magnetic flux operator of the theory, one would expect that it should transform under $S$ into the corresponding electric flux operator 
$J(S)\sim \int_S \tr\,(\phi *F)$ of the dual theory. However, this operator 
(or any appropriate generalization thereof) does not give rise to 
topological invariants, so one could question whether it should play any 
role at all. Likewise, one would like to understand the origin of the 
 shift $\left\langle {\cal O}\right\rangle\sharp 
(S\cap S)$ in the transformation of $\left\langle I(S)I(S)\right\rangle$.   

These a priori puzzling behaviours are quite  
natural when analysed from the viewpoint of Abelian electric-magnetic 
duality. In fact, 
there exists a simple Abelian topological model whose correlation 
functions mimic the behaviour in \bizarre\ under electric-magnetic duality. 
  
This model contains an Abelian gauge field $A$, whose field 
strength is defined as $F=dA$, two neutral scalar fields $\phi$, $\lambda$, 
a Grassmann-odd neutral one-form $\psi$ and a Grassmann-odd neutral two-form 
$\chi$. The Lagrangian is simply the topological density 
$$
{i\over4\pi}\tau_{0} F\wedge F ={i\tau_{0}\over 16\pi}\epsilon_{\alpha\beta\mu\nu} 
F^{\alpha\beta}F^{\mu\nu},
\eqn\chern
$$
plus conventional kinetic terms for the rest of the fields:
$$
{\cal L}_0 = {i\over4\pi}\tau_{0} F\wedge F + \im\,\tau_0\, d\phi\wedge * 
d\lambda + \im\,\tau_0\, \chi\wedge * d\psi.
\eqn\alvarito 
$$
This Lagrangian possesses the following BRST symmetry:
$$
[Q,A]=\psi,\quad \{Q, \psi\}=d\phi,\quad [Q,\phi]=[Q,\lambda]=\{Q,\chi\}=0.
\eqn\nanette
$$
Notice that the non-holomorphic metric-dependent dependence on $\tau_0$ in \alvarito\ is BRST-exact:
$$ 
\im\,\tau_0 \left ( d\phi\wedge * 
d\lambda + \chi\wedge * d\psi\right)= \im\,\tau_0 \{Q,\psi\wedge * d\lambda 
-\chi\wedge * F\}.
\eqn\zagier
$$  
Therefore, the partition function $\int{\cal D}[A,\phi,\lambda,\phi,\chi,]\ex^{\int {\cal L}_0}$ is metric-independent and 
a priori holomorphic in $\tau_0$. 

The presence of magnetic sources in the theory is mimicked by imposing the conditions: 
$$ 
\int_{S_a}F = 2\pi n^{a},\qquad n^{a}\in\ZZ, 
\eqn\unochern
$$
where the $\{S_a\}_{a=1,\cdots b_2(X)}$ are two-cycles representing a basis 
of $H_2(X;\RR)$. Notice that indeed $\int_{S}F$ gives the magnetic flux of 
$F$ through $S$. Owing to \unochern, $F$ can be decomposed as $F=da+ 
2\pi\sum_a  n^{a}[S_a]$, where $a$ is a one-form in $X$ and $[S_a]$ are 
closed two-forms representing a basis of $H^2(X;\RR)$ dual to $\{S_a\}$. 
With these conventions, the piece in $\int {\cal L}_{0}$ containing the 
field strength is simply 
$$
i\pi\tau_0\sum_{a,b}n^a Q_{ab}n^b,
\eqn\accion
$$
with $Q_{ab}=\int_X [S_a]\wedge[S_b]=\sharp(S_a\cap S_b)$ the intersection 
form of the manifold. The functional integral $\int{\cal D}A\,
\ex^{{\cal L}_0}$ therefore involves a continuous integration over 
$a$ plus a discrete summation over the magnetic fluxes $n^a$. 

We wish to calculate the correlation functions $\langle\phi^2\rangle$ and 
$\langle \int_S (2\phi F +\psi\wedge\psi)\rangle$, and analyse their 
behaviour under 
duality transformations. For this we consider the generating functional: 
$$
\int {\cal D}A{\cal D}\phi{\cal D}\lambda{\cal D}\psi{\cal D}\chi\, 
\ex^{\int_X {\cal L}(p,S)},
\eqn\odette
$$
where
$$ 
\eqalign{
{\cal L}(p,S)={i\over4\pi}\tau_0 F\wedge F &+ \im\,\tau_0\, d\phi\wedge * 
d\lambda + \im\,\tau_0\, \chi\wedge * d\psi \cr &+{1\over8\pi^2}\left(2\phi F  +\psi\wedge\psi\right)\wedge [S]+
{p\over8\pi^2}\phi^2.\cr
}
\eqn\castor
$$

Notice that the operators $\phi^2$ and $\int_S (2\phi F +\psi\wedge\psi)$ are 
BRST-invariant. This guarantees, in the usual fashion, the topological invariance of the generating function \odette. 
\REF\verlinde{E. Verlinde, ``Global Aspects of Electric-Magnetic 
Duality", {\sl Nucl. Phys.} {\bf B455} (1995), 211; hep-th/9506011.}
\REF\newresults{G. Thompson, ``New Results in Topological 
Field Theory and Abelian Gauge Theory", hep-th/9511038.}

It is possible to rewrite \odette\ in terms of an equivalent, 
dual theory, which is also a topological Abelian model of the same sort, but 
with an inverted coupling constant $\tau^{D}_{0}=-1/\tau_{0}$. The details 
are similar to those in conventional Maxwell 
electric-magnetic duality [\sdual,\verlinde,\newresults]: one introduces 
a Lagrange multiplier $A_{D}$ to enforce the constraint $F=dA$ in the functional 
integral. This $A_{D}$ can be thought of as a connection on a dual magnetic line bundle. Its field strength $F_{D}= dA_{D}$ is taken to have quantized fluxes through the cycles $S_a$: $F_{D}= da_{D}+2\pi\sum_a m^a [S_a]$. To deal with the other fields, we augment the quintet $\{F,\psi,\phi,\lambda,\chi\}$ 
with a dual Abelian field strength $F_D$, 
a dual neutral Grassmann-odd one-form $\psi_{D}$, dual 
neutral scalars $\phi_{D}$, $\lambda_D$, a dual neutral Grassmann-odd 
two-form $\chi_D$, bosonic four-form multipliers $b$, $\tilde b$, a Grassmann-odd three-form multiplier $\rho$ and a Grassmann-odd two-form multiplier 
$\omega$, and consider the extended Lagrangian:
$$
\eqalign{
\tilde{\cal L}(p,S)&={i\over4\pi}\tau_0 F\wedge F +
{\im\,\tau_0\over2\tau_0} d\phi_D\wedge * d\lambda + 
{\im\,\tau_0\over2\bar\tau_0} d\phi\wedge * d\lambda_D\cr &+ 
{\im\,\tau_0\over2\tau_0}\chi\wedge * d\psi_D  + 
{\im\,\tau_0\over2\bar\tau_0}\chi_D\wedge * d\psi \cr 
&+ {1\over8\pi^2}\left(
2\phi F +{1\over\tau_0}\psi\wedge\psi_{D}\right)\wedge [S]+
{p\over8\pi^2\tau_0}\phi\phi_{D}\cr &-{i\over2\pi}F\wedge F_{D}
+ b(\phi_{D}-\tau_0\phi) + \tilde b(\lambda_{D}-\bar\tau_0\lambda)\cr&+\rho\wedge(\psi_{D}-\tau_0\psi) 
+\omega\wedge(\chi_{D}-\bar\tau_0
\chi).\cr
}
\eqn\bellatrix
$$
By integrating out the dual fields $F_{D}$, $\phi_D$, $\psi_{D}$, $\chi_D$ 
and $\lambda_D$, and the multipliers $b$, $\tilde b$, $\rho$, $\omega$, it 
is straightforward to verify that the generating functional:
$$
\int {\cal D}[F,\phi,\psi,\lambda,\chi,F_{D},\phi_{D},\psi_{D},\lambda_D,
\chi_D,b,\tilde b,\rho,\omega]\, 
\ex^{\int_X \tilde{\cal L}(p,S)},
\eqn\capella
$$
where now the integration over $F$ is unrestricted, represents the same 
theory as \odette . The dual formulation can be 
achieved by integrating out instead the original fields $F$, $\phi$, 
$\lambda$, $\chi$ and $\psi$, together with the multipliers $b$, $\tilde b$, $\rho$ and $\omega$. One obtains in this way 
the dual Lagrangian:
$$
\eqalign{
{\cal L}_{D}(p,S)&=-{i\over4\pi\tau_0} F_D\wedge F_D +
\im\,\tau^D_0 d\phi_D\wedge * d\lambda_D + \im\,\tau^D_0 \chi_D\wedge * 
d\psi_D \cr +{1\over\tau_0{}^2}&{1\over8\pi^2}\big(2\phi_{D} F_{D}+
\psi_{D}\wedge\psi_{D}\big)\wedge [S]+
{p\over8\pi^2\tau_0{}^2}(\phi_{D})^2+{i\over2\pi\tau_0{}^3}
{(\phi_{D})^2\over8\pi^2} [S]\wedge [S].
\cr}
\eqn\pollux
$$
Notice that this dual Lagrangian is invariant under the appropriate 
dualized version of \nanette:
$$
[Q,A_D]=\psi_D,\quad \{Q, \psi_D\}=d\phi_D,\quad [Q,\phi_D]=[Q,\lambda_D]=\{Q,\chi_D\}=0.
\eqn\dualnanette
$$

From \pollux\ we can immediately read off the behaviour of the correlation 
functions under $\tau_0\to-1/\tau_0$:
$$
\eqalign{
\left\langle{1\over8\pi^2}\phi^2\right\rangle_{\tau_0}&=
\left\langle {\cal O}\right\rangle_{\tau_0}=
{1\over\tau_0{}^2}\left\langle{1\over8\pi^2}(\phi_{D})^2
\right\rangle_{-1/\tau_0}=
{1\over\tau_0{}^2}\left\langle {\cal O}\right\rangle^{D}_{-1/\tau_0},
\cr\cr
\left\langle{1\over8\pi^2}\int_S\left(2\phi F + \psi\wedge\psi\right)\right\rangle_{\tau_0}&=
\left\langle I(S)\right\rangle_{\tau_0}={1\over\tau_0{}^2} 
\left\langle{1\over8\pi^2}\int_S\left(2\phi_{D} F_{D} + \psi_{D}
\wedge\psi_{D}\right)\right\rangle_{-1/\tau_0}\cr
&={1\over\tau_0{}^2}\left\langle I(S)\right\rangle^{D}_{-1/\tau_0},
\cr\cr
\left\langle I(S)I(S)\right\rangle_{\tau_0}&= 
{1\over\tau_0{}^4}\left\langle I(S)I(S)\right\rangle^{D}_{-1/\tau_0}+{i\over2\pi\tau_0{}^3}
\left\langle {\cal O}\right\rangle^{D}_{-1/\tau_0}\sharp 
(S\cap S),\cr
}
\eqn\antares
$$
which, as promised, faithfully reproduces the modular properties \bizarre\ of the corresponding topological correlation functions of the full, non-Abelian 
theory.

\endpage

\chapter{More properties of the generating function}

In this section we will analyse more properties of the generating function \formulauno. First, we will study its behaviour in the massless limit 
$m\to 0$, then we will show that the Donaldson invariants are contained in the partition function by studying the $\cn=2$ limit ($m\to\infty$) and, finally, we will show that on $K3$ it reduces to the one obtained by Vafa and Witten 
for another twist of the $\cn=4$ supersymmetric theory.

\section{Massless limit}

We wish to analyse the behaviour of \formulauno\ and \formulados\ in 
the limit $m\to 0$. We would like to point out that there is no reason 
why this limit should make sense at all. In the massless limit, the three singularities of the low-energy description of the physical $\cn=4$ theory collapse to a unique singularity 
at $u=0$. This new singularity is a super-conformal point, whose contribution 
to the topological generating function is not straightforward to analyse 
-- see [\highrank] for a related analysis -- and need not be given by the 
naive limit of the contributions of the three singularities of the mass-deformed theory. What we find is that this limit is meaningful provided that $2\chi+3\sigma\geq 0$, but the corresponding expressions are non-trivial only when $2\chi+3\sigma = 0$. 
When $2\chi+3\sigma<0$, however, the partition function diverges as 
$m^{-3\vert 2\chi+3\sigma\vert/8}$, but it is still 
possible to extract finite predictions for certain correlation functions. 
It would be interesting to compare these results with explicit computations 
from the twisted super-conformal $\cn=4$ theory at $u=0$.     

Therefore, there are three different possible behaviours to 
be considered, depending on whether $2\chi +3\sigma$ is positive, negative 
or zero. If $2\chi +3\sigma$ is positive, both the generating function 
and the partition function vanish in the massless limit. This can be 
understood as follows: the twisted (massless) theory has the anomaly 
$-3(2\chi +3\sigma)/4$, which is strictly negative if 
$2\chi +3\sigma>0$; as the observables all have positive ghost number, 
there is no way to soak up the unbalanced fermion zero modes 
(whose net number is given by minus the anomaly) in the functional 
measure, and therefore the generating function -- and the partition 
function -- vanishes. If, on the other hand, $2\chi +3\sigma$ is 
negative, the situation is the following. Consider a certain 
correlation function $\langle{\cal O}^{(1)}\cdots {\cal O}^{(r)}\rangle$. 
The observable insertions in \formulauno\ all carry an explicit mass 
dependence dictated by their ghost number, in such a way that:  
$$
\langle{\cal O}^{(1)}\cdots {\cal O}^{(r)}\rangle\propto 
m^{3(2\chi+3\sigma)/8+\sum_{n=1}^{r}g_{n}/2},
\eqn\lata
$$
with $g_{n}$ the ghost number of the observable ${\cal O}^{(n)}$. If 
$3(2\chi+3\sigma)/8+\sum_{n=1}^{r}g_{n}/2<0$, the corresponding 
correlation function diverges in the massless limit. If, on the 
other hand, $3(2\chi+3\sigma)/8+\sum_{n=1}^{r}g_{n}/2>0$, the 
correlation function vanishes as $m\to 0$. Finally, if $3(2\chi+3\sigma)/8+\sum_{n=1}^{r}g_{n}/2=0$, which is just the anomaly-matching condition for the massless theory, the correlator is 
perfectly finite and -- a priori -- non-trivial in the massless limit.     

So as to complete the discussion, we have to study the case in which $2\chi +3\sigma=0$. In this situation the partition function is independent 
of $m$ (and therefore non-vanishing in the $m\to 0$ limit), but all the 
correlation functions vanish in this limit. This is consistent with the    
anomaly-matching condition, since when $2\chi +3\sigma=0$ the massless 
theory is anomaly-free and therefore any correlation function involving observables with non-zero ghost number must vanish.

\section{The $\cn=2$ limit and the Donaldson-Witten invariants}

We would like to analyse the fate of our formulas for the generating 
function under the decoupling limit $m\to\infty$, 
$q_0\to 0$, holding $\Lambda_{0}$, the scale of the $N_f=0$ theory, fixed: $4m^4q_0=\Lambda_{0}{}^4$. In this limit, the singularities 
at strong coupling evolve to the singularities of the $N_f=0$ $SU(2)$ 
theory, while the semiclassical singularity goes to infinity and 
disappears. While this limit is perfectly well-defined for the 
Seiberg-Witten curve, it is not clear whether the corresponding 
explicit expressions for the prepotentials and the periods should 
remain non-singular as well. In fact, one would be tempted to think that 
this is not the case, since this naive decoupling limit is highly 
singular as far as quantities such as the effective action are concerned. 
The question therefore arises as to whether taking this naive limit in the twisted theory could give a non-singular result, that is whether, 
starting from \formulauno\ or \formulados , one could recover the 
corresponding expressions for the twisted (pure) $SU(2)$ $\cn=2$ 
supersymmetric theory. This limit has been studied by Dijkgraaf et al. [\coreatres] for 
the Vafa-Witten partition function, and they were able to single out 
a piece which corresponds to the partition function of the twisted $\cn=2$ 
supersymmetric theory as first computed by Witten [\monop]. We will go 
a step further and 
recover, in the same limit, the full generating function for the Donaldson-Witten invariants.

We will focus on the generating function \formulauno . We will keep the 
leading terms in the series expansion of the different modular functions 
in powers of $q_{0}$. We will use the explicit formulas:
$$
\eqalign{
G(q_0{}^2)&=1/q_0{}^2+\cdots,\cr
G(q_0{}^{1/2})&=1/q_0{}^{1/2}+\cdots,\cr
G(-q_0{}^{1/2})&=-1/q_0{}^{1/2}+\cdots,\cr}
\qquad
\eqalign{
\vartheta_3(q_0)\vartheta_4(q_0)&=1+\cdots,\cr 
\vartheta_2(q_0)\vartheta_3(q_0)/2&=q_0{}^{1/8}+\cdots,\cr 
\vartheta_2(q_0)\vartheta_4(q_0)/2&=q_0{}^{1/8}+\cdots.\cr
}
\eqn\buchenwald
$$ 
As for the modular functions entering the observables, we have the 
following behaviour:
$$
\eqalign{
z_1={1\over4}m^2 e_1(\tau_0)&= {1\over6}m^2 + O(\Lambda_0{}^4/m^2),\cr
z_2={1\over4}m^2 e_2(\tau_0)&= -{1\over12}m^2 -4\Lambda_0{}^2+ O(\Lambda_0{}^4/m^2),\cr
z_3={1\over4}m^2 e_3(\tau_0)&= -{1\over12}m^2 +4\Lambda_0{}^2+ O(\Lambda_0{}^4/m^2),\cr
}
\eqn\lore
$$
$$
\eqalign{
(dz/da)_1{}^2&=\half m^2(\vartheta_3\vartheta_4)^4= 
\half m^2+ O(\Lambda_0{}^4/m^2),\cr
(dz/da_D)_2{}^2&=\half m^2(\vartheta_2\vartheta_3)^4= 
16 \Lambda_0{}^2+ O(\Lambda_0{}^4/m^2),\cr
(dz/da_d)_3{}^2&=-\half m^2(\vartheta_2\vartheta_4)^4= 
-16 \Lambda_0{}^2+ O(\Lambda_0{}^4/m^2),\cr
}
\eqn\decameron
$$
and, for the contact terms $T_i$ \contralto, \mezzosoprano:
$$
T_1= O(\Lambda_0{}^4/m^2),\quad 
T_2= -2\Lambda_0{}^2 + O(\Lambda_0{}^4/m^2),\quad 
T_3= 2\Lambda_0{}^2 + O(\Lambda_0{}^4/m^2).
\eqn\hades
$$

While the contribution from the semiclassical singularity behaves  
as 
$$ 
2^{-\nu}m^{3(2\chi+3\sigma)/8}q_0{}^{-\nu}\ex^{p m^2/3}\ldots,
\eqn\quark
$$
the contributions from the strong coupling singularities give 
the following result:
$$
\eqalign{
&{1\over2^{b_1}} m^{3(2\chi+3\sigma)/8}  q_0{}^{-\nu}(-1)^{\sigma/8} q_0{}^{3\nu/4}
q_0{}^{(2\chi+3\sigma)/16}\ex^{- p m^2/6}\cr&\left[
2^{1+ {1\over4}(7\chi +11\sigma)}\left(
\ex^{2p+{S^2\over2}}
\sum_{x}(-1)^{{x\over2}\cdot\xi}\ex^{S\cdot x}\,n_x 
+i^{\nu-\xi^2} \ex^{-2p-{S^2\over2}}
\sum_{x}(-1)^{{x\over2}\cdot\xi}\ex^{-iS\cdot x}\,n_x 
\right)\right],\cr
}
\eqn\donaldwitten
$$
(we have set $\Lambda_0{}^2=-1/4$), 
where the quantity in brackets is precisely Witten's generating  
function for the twisted $\cn=2$ $SO(3)$ gauge theory!

\section{Partition function on $K3$}
We now specialize to $K3$. This is a compact hyper-K\"ahler manifold, 
and as such one would expect [\wijmp] the physical and the twisted 
theories to coincide. Therefore, our formulas are to be considered 
as true predictions for the partition function and a selected set of correlation functions of the physical $\cn=4$ $SO(3)$ gauge theory on 
$K3$. 

Only the zero class $x=0$ contributes on $K3$, and $n_{x=0}=1$. 
Moreover, $\chi=24$ 
and $\sigma=-16$, so $\nu=2$ and $2\chi+3\sigma=0$. Notice that 
because of the latter identity, the mass parameter drops out from 
the formula on $K3$, a nice check. The answer for $K3$ is therefore:  
$$
Z_\xi^{K3} = {G(q_0{}^2)\over4}\delta_{\xi,0}
+ {G(q_0{}^{1/2})\over2}
+ i^{-\xi^2}{G(-q_0{}^{1/2})\over2},
\eqn\formol
$$ 
which happily coincides with the formula given by Vafa and Witten 
[\vafa]. We can go even further and present the full generating 
function on $K3$:
$$ 
\eqalign{
\left\langle\ex^{p{\cal O}+I(S)}\right\rangle^{K3}_{\xi}&=\cr   
{G(q_0{}^2)\over4}\ex^{2pz_1 + S^2 T_1}\,\delta_{\xi,0}&
+ {G(q_0{}^{1/2})\over2}\ex^{2pz_2 + S^2 T_2}
+ i^{-\xi^2}{G(-q_0{}^{1/2})\over2}\ex^{2pz_3 + S^2 T_3}.
\cr}
\eqn\formaldehido
$$
Notice that the correlation functions, which follow from \formaldehido, 
are proportional to the mass $m$, and therefore all vanish (except for 
the partition function) when $m\to 0$, as expected.

The generating function for $SU(2)$ is obtained from 
\formaldehido\ by simply setting $\xi=0$ and dividing by $2$:
$$ 
\eqalign{
\left\langle\ex^{p{\cal O}+I(S)}\right\rangle^{K3}_{SU(2)}&=\cr   
{G(q_0{}^2)\over8}\,\ex^{2pz_1 + S^2 T_1}&
+ {G(q_0{}^{1/2})\over4}\,\ex^{2pz_2 + S^2 T_2}
+ {G(-q_0{}^{1/2})\over4}\,\ex^{2pz_3 + S^2 T_3}.
\cr}
\eqn\aldehido
$$

The corresponding expression for $SO(3)$ bundles is given by the 
sum of \formaldehido\ over all 't Hooft fluxes. As explained in 
[\vafa], the allowed 't Hooft fluxes on $K3$ can be grouped into 
different diffeomorphism classes, which are classified by the value 
of $\xi^2$ modulo $4$ ($K3$ is spin, so $\xi^2$ is always even). 
There are three different possibilities and, correspondingly, three  
different generating functions to be computed: $\xi=0$, with 
multiplicity $n_0=1$, gives just the $SU(2)$ partition function; 
$\xi\not=0, \xi^2\in 4\ZZ$, with multiplicity $n_{\rm even}=
(2^{22}+2^{11})/2 -1$; and $\xi^2=2~\mod~4$, with multiplicity 
$n_{\rm odd}=(2^{22}-2^{11})/2$. The $SO(3)$ answer is the sum 
of the three generating functions (with the appropriate multiplicities):
$$ 
\eqalign{
\left\langle\ex^{p{\cal O}+I(S)}\right\rangle^{K3}_{SO(3)}&=
\left\langle\ex^{p{\cal O}+I(S)}\right\rangle^{K3}_{\xi=0}+
n_{\rm even}
\left\langle\ex^{p{\cal O}+I(S)}\right\rangle^{K3}_{\rm even}+
n_{\rm odd}
\left\langle\ex^{p{\cal O}+I(S)}\right\rangle^{K3}_{\rm odd}\cr =  
{1\over4}G(q_0{}^2)&\ex^{2pz_1 + S^2 T_1}
+ 2^{21}G(q_0{}^{1/2})\ex^{2pz_2 + S^2 T_2}
+ 2^{10}G(-q_0{}^{1/2})\ex^{2pz_3 + S^2 T_3}.
\cr}
\eqn\smirnoff
$$

All these generating functions behave under duality as dictated by 
\tduality, \thooftrule\ and \bizarre. In particular, the $S$ transformation 
exchanges the $SU(2)$ and $SO(3)$ generating functions according to 
\tdual\ and \bizarre. 

\endpage

\chapter{Conclusions and final remarks}

In this paper we have computed the generating function of all the topological correlation functions of one of the twisted $\cn =4$ supersymmetric theories,  perturbed by a mass term, for simply-connected four-manifolds of simple type. The result provides, for the first time, a framework to analyse duality transformations of correlation functions in a theory in which the duality symmetry holds exactly. The transformation properties of the topological correlation functions which have been obtained turn out to be very simple. Actually, these transformation seem to belong to a general class of duality transformations. They are the same as those of a simple Abelian topological model, in which the duality transformation can be carried out explicitly by performing standard manipulations in its functional integral formulation.

In obtaining the final expressions for the generating function, we made the ansatz \ansatz\ for the unknown function $f(m,\chi,\sigma,\tau_0)$ in the measure \measure. This ansatz is consistent with some natural properties, which one expects for the generating function. However, other choices are possible. We assumed that the partition function should transform under duality as a modular form with weight $-\chi/2$, as is the case for the theory considered by Vafa and Witten. A different hypothesis would have led to a different exponent for $\eta(\tau_0)$ in \ansatz. For the mass dependence on \ansatz, we can be confident that it is correct, as it is  the only possibility which leads to the structure dictated by the anomaly. Finally, the numerical factor could also be modified and still obtain the same results for $K3$ (as long as it remains the same when $2\chi + 3 \sigma=0$), and an equivalent behaviour in the $\cn=2$ limit. Thus, in the main result \formulauno, a global factor  
involving a modular form $g(\tau_0,\chi,\sigma)$ such that 
$g(\tau_0,\chi,-2\chi/3)=1$, could be missing. We believe, however, that our ansatz is correct but, certainly, it would be reassuring to have an independent argument to fix global factors in the $u$-plane approach, which could justify our choice.

We have restricted our analysis to the case of simply-connected manifolds of simple type. The generalization of \formulauno\ to the non-simple type case 
can be easily done within the framework of the $u$-plane approach, by using the general formula $\sw$. It would be very interesting to find out if the modular properties found in the simple-type case hold in general or, conversely, if they do not hold, which implications can be inferred for the higher-order Seiberg-Witten invariants.

Our work shows, following [\wimoore,\highrank,\nonsimply,\strings], how wall-crossing techniques within the $u$-plane approach can be implemented to obtain explicit expressions for topological quantities. It would be very interesting to study if the same methods can be applied to the twist considered by Vafa and Witten to reobtain their results, and to extend them to more general types of manifolds. Another important issue that should be considered is the extension of our results, as well as those obtained by Vafa and Witten, to the case of $SU(N)$ for arbitrary $N$. We expect to address some of these 
issues in future work.

\vskip2cm
\ack
We would like to thank L. Alvarez-Gaum\'e and J.L.F. Barb\'on for helpful discussions, and M. Mari\~no for  
many useful remarks and for making available to us his results 
on $\cn=4$ supersymmetric theories, and for a careful reading of the 
manuscript. One of us (C.L.), wishes to thank the Theory Division at 
CERN for its hospitality. 
This work was supported in part by DGICYT under grant PB96-0960
and by the EU Commission under TMR grant FMAX-CT96-0012.

\endpage

%%%%%%%%%%%%%%%%%%%%%%%%%%%%%%%%%%%%%%%%%%%%%%%%%%%%%%%%%%%%%%%%%%
%%                                                              %%
%%%%%%%%%%%%%%%%%%%%%%%%%%%%%%%%%%%%%%%%%%%%%%%%%%%%%%%%%%%%%%%%%%
\appendix 

Here we collect some useful formulas which should help the reader 
follow the computations in the paper. A more detailed 
account can be found in appendices A and B of [\wimoore]. A very useful  
review containing definitions and properties of many modular forms 
can be found in appendices A and F of 
\REF\kiritsis{E. Kiritsis, ``Introduction to Superstring Theory", 
hep-th/9709062.}[\kiritsis].

\section{Modular forms}

Our conventions for the Jacobi theta functions are:
$$
\eqalign{
\vartheta _2 (\tau ) &= \sum _{n=-\infty}^{\infty} 
q^{(n+1/2)^2/2}= 2q^{1/8}(1+ q + q^3 + \cdots),\cr 
\vartheta _3 (\tau ) &= \sum _{n=-\infty}^{\infty} q^{n^2/2}
= 1 + 2 q^{1/2} + 2 q^2 + \cdots,\cr
\vartheta _4 (\tau ) &= \sum _{n=-\infty}^{\infty} (-1)^n q^{n^2/2}
= 1 - 2 q^{1/2} + 2 q^2 +\cdots,\cr}
\eqn\jacobi
$$
where $q=\ex^{2\pi i\tau}$. 
They satisfy the identity:
$$
\vartheta _3 (\tau )^4=\vartheta _2 (\tau )^4+\vartheta _4 (\tau )^4.
\eqn\cris
$$

They have the following properties under modular transformations:
$$
\eqalign{
\vartheta _2 (-1/\tau ) &= \sqrt{\tau\over i}\vartheta _4 (\tau ),\cr
\vartheta _3 (-1/\tau ) &= \sqrt{\tau\over i}\vartheta _3 (\tau ),\cr
\vartheta _4 (-1/\tau ) &= \sqrt{\tau\over i}\vartheta _2 (\tau ),
\cr}\qquad\qquad
\eqalign{
\vartheta _2 (\tau+1 ) &= \ex^{i\pi/4}\vartheta _2 (\tau ),\cr
\vartheta _3 (\tau+1 ) &= \vartheta _4 (\tau ),\cr
\vartheta _4 (\tau+1 ) &= \vartheta _3 (\tau ).
\cr}
\eqn\jacobino
$$
 
From these, the modular properties of the functions 
$e_j$ \spin\ follow straightforwardly:
$$
\eqalign{
e_1(-1/\tau_0)&= \tau_0{}^2 e_2(\tau_0),\cr
e_2(-1/\tau_0)&= \tau_0{}^2 e_1(\tau_0),\cr
e_3(-1/\tau_0)&= \tau_0{}^2 e_3(\tau_0),\cr
}\qquad\qquad
\eqalign{
e_1(\tau_0+1)&= e_1(\tau_0),\cr
e_2(\tau_0+1)&= e_3(\tau_0),\cr
e_3(\tau_0+1)&= e_2(\tau_0).\cr
}
\eqn\irredento
$$
Notice that, from their definition, $e_1+e_2+e_3=0$. 
Likewise, we can determine explicitly the behaviour of the 
functions $\kappa_{j}$ \yespadas\ and of the periods 
\zhivago\ under modular transformations:
$$ 
\eqalign{
\kappa_{1}(-1/\tau_0)&=\tau_0{}^2\kappa_{2}(\tau_0),\cr  
\kappa_{2}(-1/\tau_0)&=\tau_0{}^2 \kappa_{1}(\tau_0),\cr
\kappa_{3}(-1/\tau_0)&=-\tau_0{}^2 \kappa_{3}(\tau_0),
\cr}
\qquad\qquad
\eqalign{
\kappa_{1}(\tau_0+1)&=-\kappa_{1}(\tau_0),\cr
\kappa_{2}(\tau_0+1)&=\kappa_{3}(\tau_0),\cr
\kappa_{3}(\tau_0+1)&=\kappa_{2}(\tau_0),\cr
}
\eqn\calatayud
$$
and
$$ 
\eqalign{
(da/dz)_1{}^2\vert_{-{1\over\tau_0}}& =
\tau_0{}^{-4} (da_{D}/dz)_2{}^2\vert_{\tau_0},\cr  
(da_{D}/dz)_2{}^2\vert_{-{1\over\tau_0}}&= 
\tau_0{}^{-4} (da/dz)_1{}^2\vert_{\tau_0},\cr
(da_{d}/dz)_3{}^2\vert_{-{1\over\tau_0}}&= 
\tau_0{}^{-4}(da_{d}/dz)_3{}^2\vert_{\tau_0},
\cr}
\qquad
\eqalign{
(da/dz)_1{}^2\vert_{\tau_0+1}&= (da/dz)_1{}^2\vert_{\tau_0},\cr
(da_{D}/dz)_2{}^2\vert_{\tau_0+1}&= (da_{d}/dz)_3{}^2
\vert_{\tau_0},\cr
(da_{d}/dz)_3{}^2\vert_{\tau_0+1}&= (da_{D}/dz)_2{}^2
\vert_{\tau_0},\cr
}
\eqn\calahorra
$$
where we have set $a_{d}\equiv a_{D}-a$.

The Dedekind eta function is defined as follows:
$$
\eta(\tau)=q^{1/24}\prod^{\infty}_{n=1}(1-q^n)=\sum_{-\infty}
^{\infty}(-1)^{n}q^{{3\over2}(n-1/6)^2}= q^{1/24}(1 - q - q^2 + 
\cdots).
\eqn\dedekind
$$
Under the modular group:
$$
\eta(-1/\tau)=\sqrt{\tau\over i}\eta(\tau),\qquad 
\eta(\tau+1)=\ex^{i\pi/12}\eta(\tau).
\eqn\borrell
$$
The following identities are useful:
$$
\eqalign{
\vartheta _2 (\tau )\vartheta _3 (\tau )\vartheta _4 (\tau )
&=2\eta(\tau)^3,\cr\cr
\vartheta _2 (\tau )= 2 {\eta(2\tau)^2\over\eta(\tau)},\qquad 
\vartheta _3 (\tau )&= {\eta(\tau)^5\over\eta(\tau/2)^2\eta(2\tau)^2},
\qquad 
\vartheta _4 (\tau )= {\eta(\tau/2)^2\over\eta(\tau)}.
\cr}
\eqn\almunia
$$
With these formulas we can rewrite the functions $G(q)$ 
featuring in 
the Vafa-Witten formula in terms of standard modular forms:
$$
\eqalign{
G(q)&= {1\over\eta(q)^{24}},\cr 
G(q^2)&= {1\over\eta(2\tau)^{24}}= \left({2\over\eta(\tau)\vartheta_2(\tau)}\right)^{12},\cr
}\quad
\eqalign{
G(q^{1/2})&= {1\over\eta(\tau/2)^{24}}= {1\over\bigl(\eta(\tau)\vartheta_4(\tau)\bigr)
^{12}},\cr
G(-q^{1/2})&= {1\over\eta({\tau+1\over2})^{24}}= -{1\over\bigl(\eta(\tau)\vartheta_3(\tau)\bigr)^{12}}.
\cr} 
\eqn\valiente 
$$
These functions have the following modular properties [\vafa]:
$$
\eqalign{
G(q^{2})&{\buildrel\tau\to\tau+1\over\too}    G(q^{2}),\cr
G(q^{1/2})&{\buildrel\tau\to\tau+1\over\too}  G(-q^{1/2}),\cr
G(-q^{1/2})&{\buildrel\tau\to\tau+1\over\too} G(q^{1/2}),
\cr}\qquad\qquad
\eqalign{
G(q^{2})&{\buildrel\tau\to -1/\tau\over\too}    2^{12}\tau^{-12} 
G(q^{1/2}),\cr
G(q^{1/2})&{\buildrel\tau\to -1/\tau\over\too}  2^{-12}\tau^{-12} 
G(q^{2}),\cr 
G(-q^{1/2})&{\buildrel\tau\to -1/\tau\over\too} \tau^{-12} 
G(-q^{1/2}).
\cr}
\eqn\conhazo
$$

The Eisenstein series of weights 2 and 4 are:
$$
\eqalign{
E_2 & = {12\over i\pi}\partial_{\tau}{\rm log}\,\eta = 
1-24\sum_{n=1}^{\infty} {n q^n\over 1-q^n}=1 - 24 q + \cdots ,\cr
E_4 & =\half\left (\vartheta_2^8+\vartheta_3^8+\vartheta_4^8\right)
=1+240\sum_{n=1}^{\infty} {n^3 q^n\over 1-q^n}.\cr}
\eqn\cazzo
$$
While $E_4$ is a modular form of weight $4$ for $Sl(2,\ZZ)$, $E_2$ is 
not quite a modular form: under $\tau\to (a\tau+b)/(c\tau+d)$ 
we have:
$$
E_2\left({a\tau+b\over c\tau+d}\right)=(c\tau+d)^2 E_2(\tau)-
{6ic\over\pi}(c\tau+d). 
\eqn\uno 
$$
The 
non-holomorphic combination $\hat E_2 = E_2 - 3/(\pi\im\,\tau)$ is a 
modular form of weight $2$, which enters in the definition of the 
contact term $\hat T$ in \karembeu . 
 
\section{Lattice sums} 
Here we quote some of the results in appendix B of [\wimoore], 
to which we refer the reader for more details. These 
formulas are quite useful when performing the duality 
transformations among the different frames on the $u$-plane. 

We introduce the following theta function:
$$
\eqalign{
\Theta_{\Gamma} (\tau, \alpha,\beta; P, \gamma)
\equiv\,
&
 \exp\left[{ \pi \over  2 y} ( \gamma_+^2 - \gamma_-^2) \right] \cr
\sum_{\lambda\in\Gamma}
\exp\biggl\{ i \pi \tau (\lambda+ \beta)_+^2 +
i \pi \bar \tau (\lambda+ \beta)_-^2
&
+ 2 \pi i (\lambda+\beta)\cdot\gamma - 2 \pi i
\left(\lambda+\half \beta\right)\cdot\alpha \biggr\} \cr
= & \,\ex^{i \pi \beta\cdot\alpha}\,
 \exp\left[{ \pi \over  2 y} ( \gamma_+^2 - \gamma_-^2)\right] \cr
\sum_{\lambda\in\Gamma}
\exp\biggl\{ i \pi \tau (\lambda+ \beta)_+^2 +
i \pi \bar \tau (\lambda+ \beta)_-^2
&
+ 2 \pi i (\lambda+\beta)\cdot\gamma - 2 \pi i
(\lambda+  \beta)\cdot\alpha \biggr\}, \cr}
\eqn\caspa
$$
where $\Gamma$ is a lattice of signature $(b_+,b_-)$ on which 
an orthogonal projection $P_\pm(\lambda)= \lambda_\pm$ and 
a pairing $(\alpha,\beta)\to \alpha\cdot\beta\in\RR$ are defined.  

The main transformation law is:
$$
\Theta_{\Gamma }\left(-1/\tau, \alpha,\beta; P, {\gamma_+ \over  \tau} +
{\gamma_- \over  \bar \tau}\right)
= \sqrt{\vert \Gamma \vert \over  \vert \Gamma^{*} \vert}
(-i \tau)^{b_+/2} (i \bar \tau)^{b_-/2}
\Theta_{\Gamma^{*}} ( \tau, \beta,-\alpha ; P, \gamma),
\eqn\lolita
$$
where $\Gamma^{*}$ is the dual lattice.
Given a characteristic vector $w_2\in\Gamma$, such that
$$
\lambda\cdot\lambda = \lambda\cdot w_2 ~ {\rm mod} ~2
\eqn\vangaal
$$
for all $\lambda\in\Gamma$,
then we have:
$$
\Theta_{\Gamma} (\tau+1, \alpha,\beta; P, \gamma)
= \ex^{-i \pi\beta\cdot w_2/2}
\Theta_{\Gamma }\left( \tau, \alpha-\beta-\half w_2,\beta ; P, \gamma\right)
\eqn\cruyff
$$

%\end

\vfill
\endpage

\refout
\end